\documentclass[12pt,a4paper]{article}
\usepackage{amsmath,caption,amsfonts,setspace,mathrsfs,mathtools,bbm}
\usepackage[top=1.2in, bottom=1.2in, left=1in, right=1in]{geometry}

\usepackage[round]{natbib}
\usepackage{color,soul}
\renewcommand{\baselinestretch}{1.4} 
\DeclareCaptionStyle{italic}[justification=centering]{labelfont={bf},textfont={it},labelsep=colon}
\usepackage{graphicx,psfrag,epsf,textcomp}
\usepackage{enumerate, dsfont}
\usepackage{natbib}
\usepackage{url,xcolor} 
\usepackage{booktabs, subfig, bm, paralist,mathpazo,tikz,longtable,microtype,array,appendix}

\newcolumntype{L}[1]{>{\raggedright\let\newline\\\arraybackslash\hspace{0pt}}m{#1}}
\newcolumntype{C}[1]{>{\centering\let\newline\\\arraybackslash\hspace{0pt}}m{#1}}
\newcolumntype{R}[1]{>{\raggedleft\let\newline\\\arraybackslash\hspace{0pt}}m{#1}}

\usepackage[pdftex,colorlinks=true]{hyperref}
\definecolor{darkblue}{rgb}{0,0,.6}
\hypersetup{citecolor=darkblue,linkcolor=darkblue,urlcolor=darkblue}

\date{}

\newcommand{\blind}{0}

\newcommand{\I}{\mathcal{I}}

\graphicspath{{plots/}}
\DeclareMathOperator*{\argmin}{\arg\!\min}
\newsavebox\CBox

\newcommand{\inp}[2]{\langle #1\, , #2 \rangle}
\definecolor{a0}{rgb}{0.0, 0.5, 0.0}
\definecolor{bistre}{rgb}{0.24, 0.17, 0.12}
\definecolor{amethyst}{rgb}{0.6, 0.4, 0.8}
\definecolor{blue-violet}{rgb}{0.54, 0.17, 0.89}
\definecolor{Rcolor}{RGB}{150,160,190}
\definecolor{blush}{rgb}{0.87, 0.36, 0.51}
\definecolor{brightturquoise}{rgb}{0.03, 0.91, 0.87}
\definecolor{burntorange}{rgb}{0.8, 0.33, 0.0}

\begin{document}

\def\spacingset#1{\renewcommand{\baselinestretch}{#1}\small\normalsize} \spacingset{1}

\if0\blind
{
  \title{\bf Forecasting Australian subnational age-specific mortality rates}
  \author{Han Lin Shang\thanks{Postal address: Department of Actuarial Studies and Business Analytics, Level 7, 4 Eastern Rd, Macquarie University, Sydney, NSW 2109, Australia; Email: hanlin.shang@mq.edu.au; ORCID: \url{https://orcid.org/0000-0003-1769-6430}.}
  \hspace{.2cm}\\
    Department of Actuarial Studies and Business Analytics \\
    Macquarie University
    \\
    \\
    Yang Yang \\
    Research School of Finance, Actuarial Studies and Statistics \\
    Australian National University}
  \maketitle
} \fi

\if1\blind
{
    \title{\bf Forecasting Australian subnational age-specific mortality rates}
    \author{}
  \maketitle
} \fi

\begin{abstract}
When modeling sub-national mortality rates, it is important to incorporate any possible correlation among sub-populations to improve forecast accuracy. Moreover, forecasts at the sub-national level should aggregate consistently across the forecasts at the national level. In this study, we apply a grouped multivariate functional time series to forecast Australian regional and remote age-specific mortality rates and reconcile forecasts in a group structure using various methods. Our proposed method compares favorably to a grouped univariate functional time series forecasting method by comparing one-step-ahead to five-step-ahead point forecast accuracy. Thus, we demonstrate that joint modeling of sub-populations with similar mortality patterns can improve point forecast accuracy. 

\vspace{0.2cm}

\end{abstract}

\vspace{.2in}



 \newpage
 \spacingset{1.56}

 \section{Introduction}\label{sec:intro}

Mortality is regarded as one of the most widely available measures of health conditions in a community. National and sub-national mortality patterns in terms of age, sex, and geographical distribution are of interest to epidemiologists, health care personnel, and those working in health and social policy, planning, and administration. Traditional studies of human mortality focus on population data at the national level. In recent years, academics and government stakeholders have been showing increasing interest in regional mortality improvements after realizing that sub-national forecasts of age-specific mortality rates are useful for informing social and economic policies within local regions. Thus, any improvement in the forecast accuracy of regional mortality rates would help determine the allocation of current and future resources at the national and sub-national levels. Many mortality modeling methods have been proposed since the publication of the Gompertz law in 1825 \citep[see][for a comprehensive literature review on mortality modeling and forecasting]{Booth2008}, but only a few approaches can forecast multiple regional mortality rates within a country simultaneously. Thus, proposing a multivariate functional time series method for producing accurate and coherent sub-national age-specific mortality forecasts is the primary motivation of our study.

Functional time series can arise by separating an almost continuous time record into consecutive intervals, such as the intraday trajectories of the S\&P 500 index \citep{Shang2017forecasting} and the monthly sea surface temperature in climatology \citep{Shang2011}. Alternatively, functional time series can also arise when the observations in a period can be considered together as finite realizations of an underlying continuous function, for example, age-specific mortality rates in demography \citep[see][]{HU07, CM09}. In either case, the functions obtained form a time series $\{f_t, t\in \mathbb{Z}\}$, where each $f_t$ is a random function of a stochastic process and $x\in \mathcal{I}$ represents a continuum bounded within a finite interval. In this study, we consider age-specific mortality rates observed annually as a functional time series whose continuum is the age of the population.

The increasing popularity of the functional time series has resulted in a rapid increase in the literature on functional time series modeling and forecasting. From a parametric perspective, \cite{Bosq00} and \cite{Bosq08} proposed the functional autoregressive of order 1 (FAR(1)). They derived one-step-ahead forecasts that are based on a regularized form of the Yule-Walker equations. Later, FAR(1) was extended to FAR($p$), under which the order $p$ can be determined via \citeauthor{KR13}'s \citeyearpar{KR13} hypothesis testing procedure. To overcome the difficulty of infinite-dimensional parameter estimation, \cite{ANH15} showed the asymptotic equivalence between a FAR and a vector autoregressive (VAR) model via a functional principal component analysis (FPCA) and proposed a forecasting method based on the VAR forecasts of principal component scores. The approach of \cite{ANH15} can also be viewed as an extension of \cite{HS09}. \cite{HS09} forecasts principal component scores by a univariate time series forecasting method. Recently, \cite{KK16} proposed the functional moving average (FMA) process and introduced an innovative algorithm to obtain the best linear predictor. \cite{KKW16} extended the VAR model to the vector autoregressive moving average (VARMA) model. The VARMA model is a simpler estimation approach of the functional autoregressive moving average model. Recently, \cite{LRS16} considered long-range dependent curve time series and proposed a functional ARIMA model. From a nonparametric perspective, \cite{BCS00} proposed the functional kernel regression method to model temporal dependence via a similarity measure characterized by semi-metric, bandwidth, and kernel functions. From a semi-parametric viewpoint, \cite{AV08} put forward a semi-functional partial linear model that combines parametric and nonparametric models. The semi-functional partial linear model allows us to consider additive covariates and to use a continuous path in the past to predict a stochastic process's future values. Apart from the estimation of a conditional mean, \cite{HHR13} consider a functional analog of the autoregressive conditional heteroskedasticity model for modeling conditional variance. In contrast, \cite{AHP17} consider a functional analog of the generalized autoregressive conditional heteroskedasticity model. \cite{KRS17} considered a portmanteau test for testing autocorrelation under a functional generalized autoregressive conditional heteroskedasticity model. 

In the functional time series literature, limited work has been conducted on jointly modeling and forecasting multiple sub-national mortality rates over some time. \cite{Shang16c} attempted to simultaneously model and forecast state-specific mortality in Australia by determining a common trend across populations and sex-specific trends of each state. Another noticeable exception is the work by \cite{Gao2019} in which an additive factor model is fitted to functional principal component scores obtained in the initial dimension reduction of smoothed sub-national mortality data. However, both approaches mentioned above fail to explicitly consider significant correlations between sub-national populations in basis function decomposition and forecasting and thus lose valuable information that can be exploited to improve forecasting accuracy. Moreover, applying existing functional time series forecasting methods to sub-national mortality disaggregated by attributes such as sex, state, or ethnicity independently does not ensure coherence in the forecasts. That is, the forecasts of all subpopulations will not add up to the forecasts obtained by applying the technique to the national data. Hence, in practice, it is important to consider reconciliation approaches \citep[see, e.g.,][]{SCM42, SW09} in forecasting sub-national mortality rates. 

In this study, we apply the univariate and multivariate functional time series forecasting methods to the Australian national and sub-national age-specific mortality rates from 1993 to 2016 and compare their point forecast accuracies. To ensure forecasts of disaggregated series consistently aggregating to the national totals, we consider a group structure for mortality series. We apply the bottom-up, optimal combination, trace minimization reconciliation methods, and their forecast combination to reconcile forecasts. Evaluated by the mean absolute forecast error and the root mean square forecast error, we find that the trace minimization and forecast combination methods have similar superior point forecast accuracy. Our empirical study results prove that jointly modeling of populations with similar mortality patterns can improve point forecast accuracy.

The rest of this report is structured as follows. In Section~\ref{sec:data}, we describe the motivating dataset, which is the Australian national and sub-national age-specific mortality rates. In Section~\ref{sec:method}, we introduce the multivariate functional principal component regression for producing point forecasts. Forecast reconciliation methods for the grouped functional time series are discussed in Section~\ref{sec:recon}. Reconciled point forecasts produced by univariate and multivariate forecasting methods are evaluated and compared in Section~\ref{sec:result}.  Conclusions are presented in Section~\ref{sec:conclusion}. 

\section{Australian age-specific mortality rates} \label{sec:data}

\subsection{Overview of data} \label{sec:2.1}

\begin{figure}[!htb]
 \centering
 \subfloat[Observed female mortality rates]
 {\includegraphics[width = 3.3in]{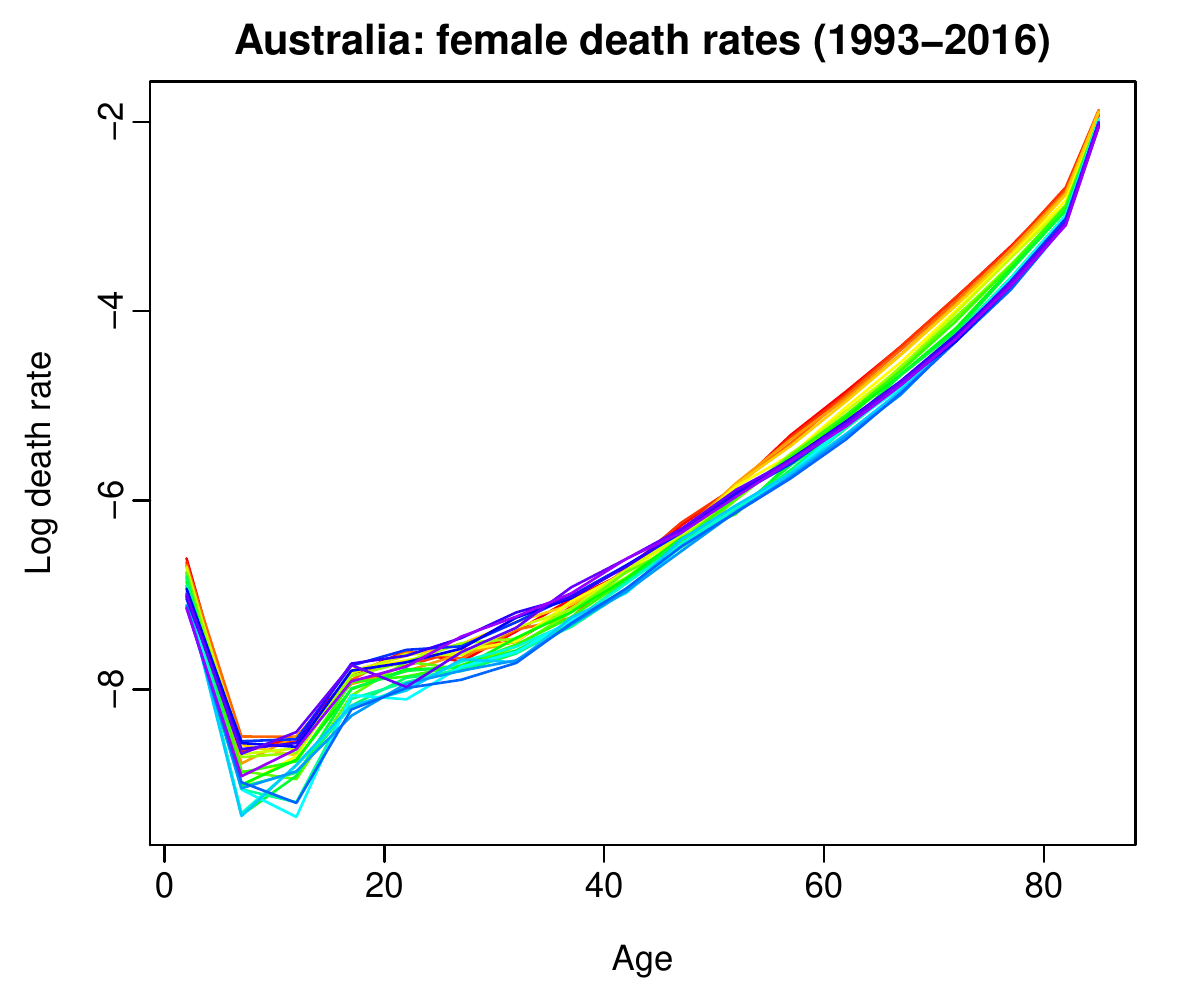}} \qquad
 \subfloat[Observed male mortality rates]
 {\includegraphics[width = 3.3in]{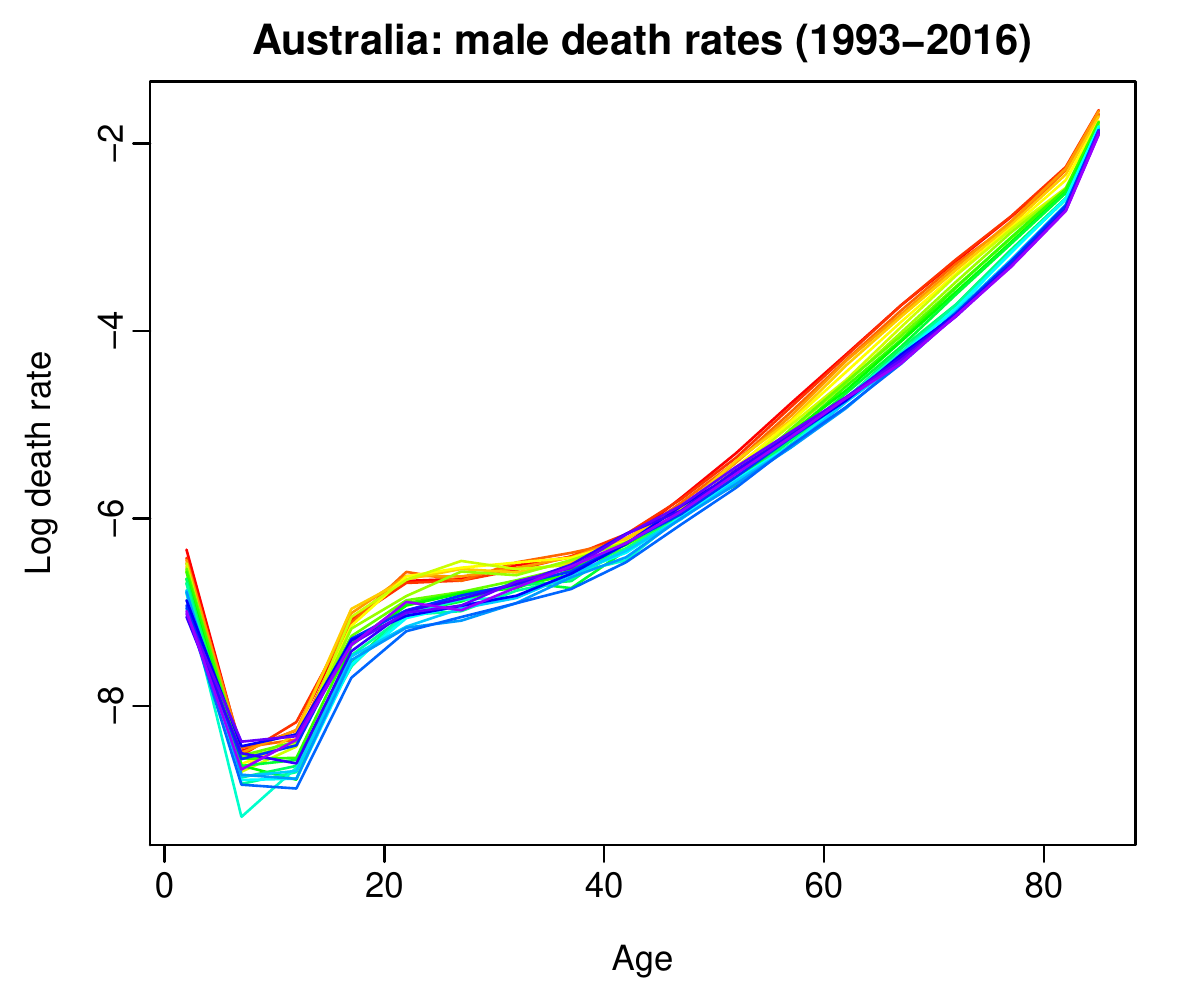}} \\
 \subfloat[Smoothed female mortality rates]
 {\includegraphics[width = 3.3in]{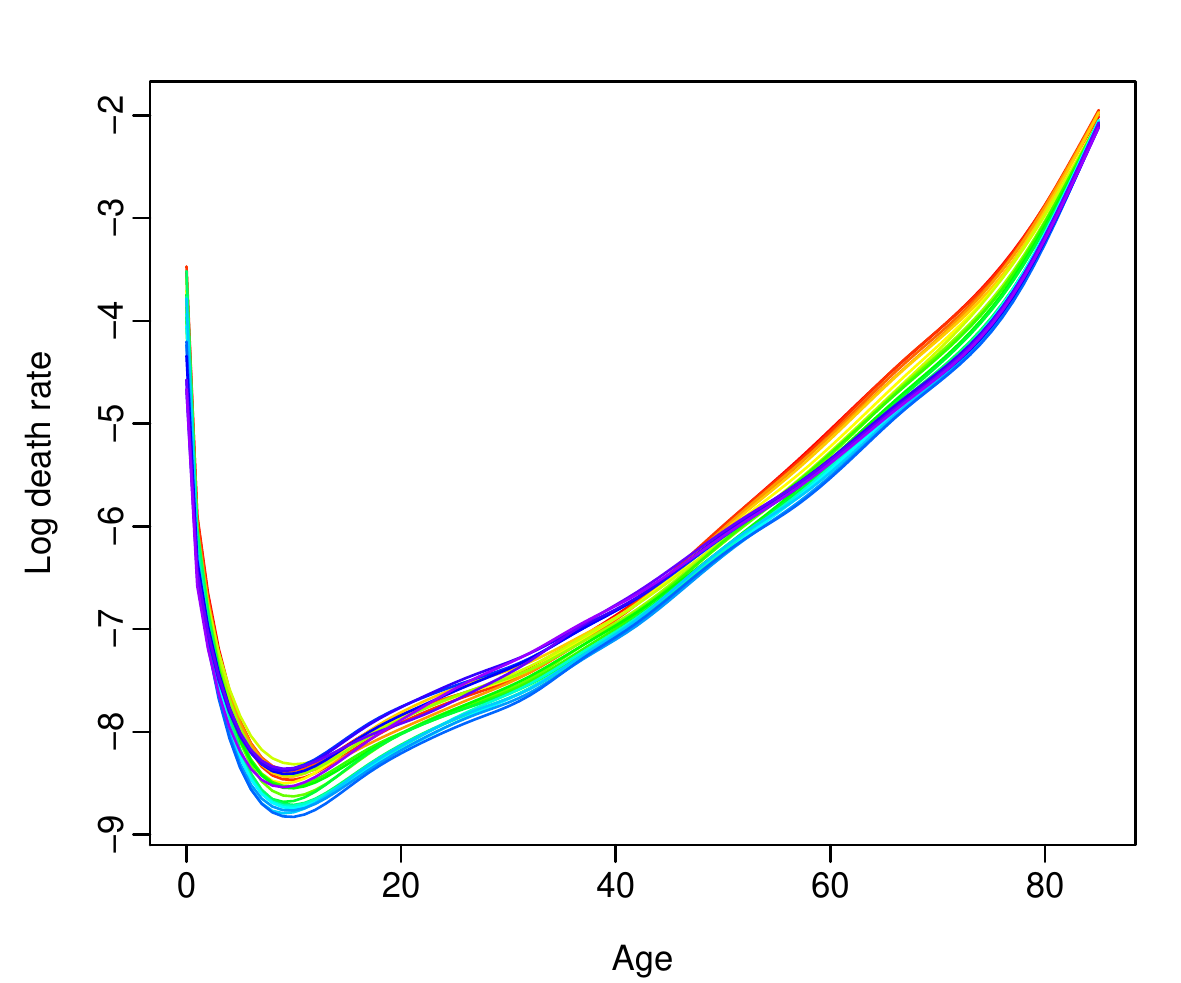}}\qquad
 \subfloat[Smoothed male mortality rates]
 {\includegraphics[width = 3.3in]{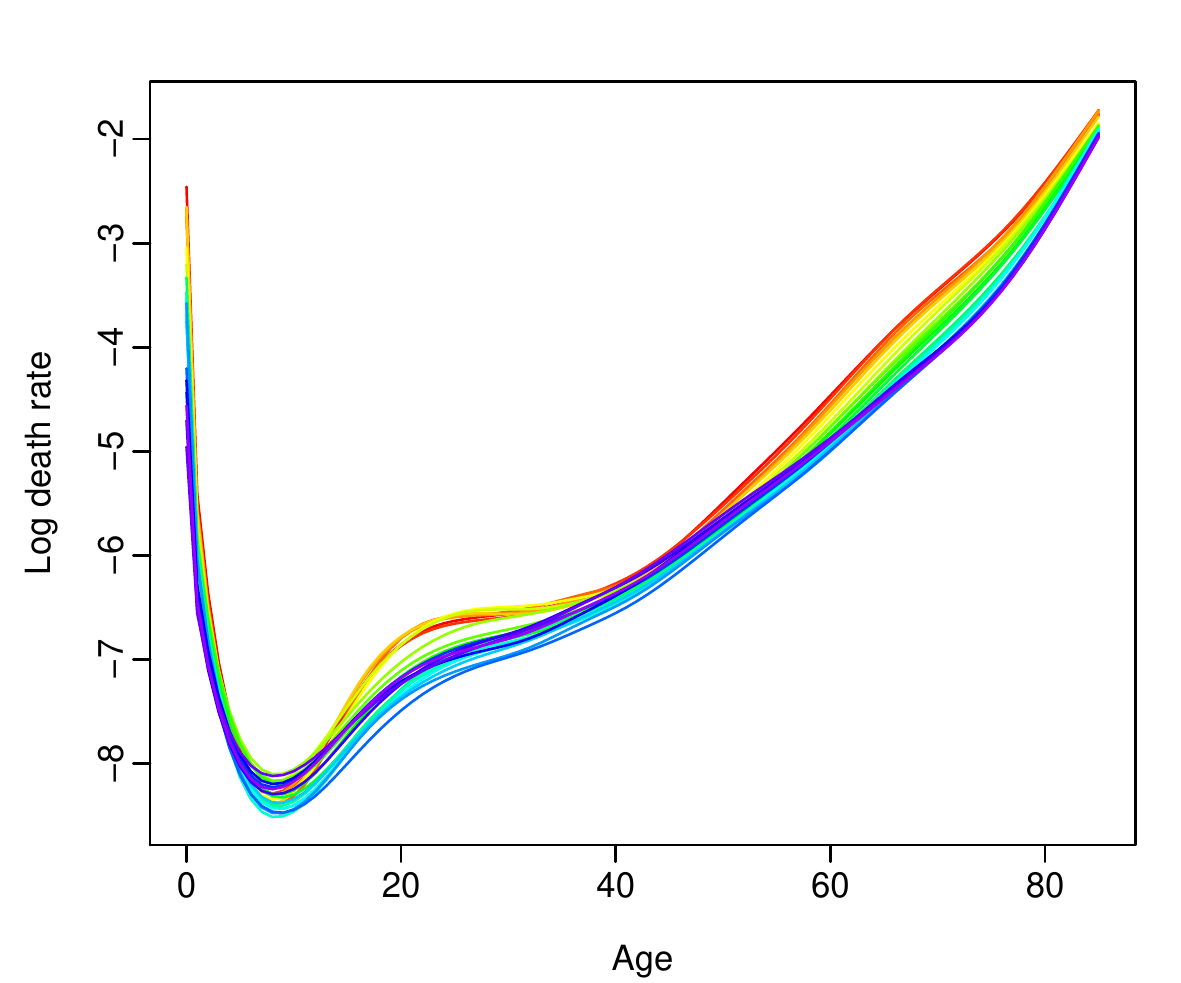}} 
 \caption{Functional time series graphical displays. The nonparametric smoothing method in~\eqref{sec:2.2} can be used to smooth age-specific mortality rates.}\label{fig: fts}
 \end{figure}

Australia, as a developed country, faces increasing longevity risks and challenges to the sustainability of the pension, health, and aged care systems caused by an aging population\citep[see, e.g.,][]{Coulmas07, OECD13}. To better understand the ongoing mortality improvements, we studied the Australian age-specific mortality rates in 1993-2016. We consider ages from 0 to 85 years in 5-year age groups and include all ages at and beyond 85 years in the last age group, thus considering 18 groups. Figure~\ref{fig: fts} presents rainbow plots of the age-specific mortality rates of females and males in the period studied. The ``J" shaped death curves shown in the figure indicate typical mortality patterns that rapidly decrease in infant ages, and a steady increase in middle-to-old age groups. There is also a noticeable ``accident hump'' during the teenage years for the Australian male population. The rainbow plot is designed to display the distant past curves in red and the more recent curves in purple \citep{HS10}, indicating an overall reduction in the trend of total mortality rates for Australian people. However, death rates for males and females aged less than 50 have a sudden increase in 2012, and then gradually fall off with time, as shown in the bottom panels in Figure~\ref{fig: fts}.

\subsection{Nonparametric smoothing} \label{sec:2.2}

The observed mortality rates contaminated by noise are shown in the top panel of Figure~\ref{fig: fts}. To obtain smooth functions and deal with possibly missing values, we consider a penalized regression spline smoothing with monotonic constraint while preserving the shape of the log mortality curves \citep[see also][]{HU07, Shang16c}. For a group of populations, let $y_t^j(x_i)$ be the log central mortality rates observed at the beginning of each year for year $t=1,2,\dots,n$ at observed ages $x_1, x_2,\dots, x_p$ where $x$ denotes the age variable, $p$ denotes the number of ages considered, and superscript $j$ represents an individual population. The value of $j$ depends on the number of populations considered. For example, $j$ is binary when representing age-specific mortality rates of females and males for a particular area (at the national or sub-national level); $j = 1, 2, \cdots, 11$ if \textit{Region} level (described in Section~\ref{sec:2.3}) total mortality rates are considered. 

We assume that there is an underlying continuous and smooth function $f_t^j(x)$ such that
\begin{equation*}
 y_t^j(x_i) = f_t^j(x_i) + \delta_t^j(x_i) \varepsilon_{t,i}^j,
\end{equation*}
where $x_i$ represents the center of each age or age group, $\{\varepsilon_{t,i}^j\}$ are independent and identically distributed random variables across $t$ and $i$ with a mean of zero and a unit variance, and $\delta_t^j(x_i)$ measures the variability in mortality at each age in year $t$ for the $j^{\text{th}}$ population. Jointly, $\delta_t^j(x_i)\varepsilon_{t,i}^j$ represents the smoothing error.

Let $m_t^j(x_i) = \exp[y_t^j(x_i)]$ be the observed central mortality rates for age $x_i$ in year $t$. Define $N_t^j(x_i)$ to be the total $j^{\text{th}}$ population of age $x_i$ at 1\textsuperscript{st} January of year $t$. The observed log mortality rate approximately follows a Poisson distribution with estimated variance
\begin{equation}
\Big(\widehat{\delta}_t^j\Big)^2(x_i) := \text{Var}\left\{\ln [m_t^j(x_i)]\right\} \approx \frac{1}{m_t^j(x_i)\times N_t^j(x_i)}. \label{eq:1}
\end{equation}

In line with an earlier study by \cite{HU07}, we smooth log mortality rates using weighted penalized regression splines with a partial monotonic constraint for ages above 65 years, where the weights are equal to the inverse variance given in~\eqref{eq:1}, that is, $w_t^j(x_i) = 1/(\widehat{\delta}_t^j)^2(x_i)$. The penalized regression spline can be written as
\begin{equation*}
\widehat{f}_t^j(x_i) = \argmin_{\theta_t(x_i)}\sum^M_{i=1}w_t^j(x_i) \Big|y_t^j(x_i) - \theta_t(x_i)\Big| + \tau \sum^{M-1}_{i=1} \Big|\theta_t^{'}(x_{i+1}) - \theta_t^{'}(x_i)\Big|,
\end{equation*}
where $i$ represents different ages (grid points) in a total of $M$ grid points, $\theta{'}$ denotes the first derivative of smooth function $\theta$, and $\tau$ is a smoothing parameter. Both $\theta{'}$ and $\theta$ can be approximated by a set of $B-$spline basis \citep[see, e.g., ][]{de2001}. While the $L_1$ loss function and the $L_1$ roughness penalty are employed to obtain robust estimates, the monotonic increasing constraint helps to reduce the noise from the estimation of older ages \citep[see also][]{HN99}. 

\subsection{Multivariate functional time series of mortality rates} \label{sec:2.3}

The vast land area of Australia can be divided into six states: New South Wales (NSW), Victoria (VIC), Queensland (QLD), South Australia (SA), Western Australia (WA), and Tasmania (TAS), as well as two internal territories, namely, the Northern Territory (NA) and the Australian Capital Territory (NT) (ACT). Based on the remoteness, areas with similar characteristics of eight considered states and territories are often grouped into geographic classifications. We follow \cite{Guan18} and classify sub-national population statistics sourced from the Australian Bureau of Statistics (ABS) into 47 harmonized geographic areas based on the Statistical Division level of ABS. In this study, we adopt a structure similar to the PRMA classification \citep{DD94}. As shown in Table~\ref{tab:1}, we classify 47 areas into Capital City\footnote{The capital city of NT, Darwin, is grouped into Regional Australia.} and four broader regions, i.e., NSW Coast, Country Victoria, Regional Australia, and Remote Australia.

\begin{table}[!ht]
	\centering
	\caption{Remoteness classification in Australia.}\label{tab:1}
	\begin{tabular}{ | C{1cm} | L{3.5cm}| C{1cm} | L{5cm} | C{1cm}| L{4.4cm} | }
		\hline
		\multicolumn{2}{|c|}{\textbf{Capital City}} & \multicolumn{2}{|c|}{\textbf{Regional Australia}} & \multicolumn{2}{|c|}{\textbf{Remote Australia}} \\
		\hline 
		Area & Name & Area & Name & Area & Name \\
		\hline
		1 & Sydney & 16 & North West NSW & 35 & West NSW \\ 
		5 & Melbourne & 17 & Central West NSW & 36 & South West Queensland \\ 
		11 & Brisbane & 18 & Murrumbidgee & 37 & Central West Queensland \\ 
		12 & Adelaide & 19 & Murray & 38 & Far North \\ 
		13 & Perth & 20 & Wimmera & 39 & North West \\ 
		14 & Greater Hobart & 21 & Mallee & 40 & Eyre \\ 
		15 & Canberra & 22 & Ovens-Murray & 41 & Northern SA \\ 
		\cline{1-2} 
		\multicolumn{2}{|c|}{\textbf{NSW Coast}} & 23 & Wide Bay-Burnett \& Fitzroy & 42 & South Eastern WA \\
		\cline{1-2}
		2 & Hunter & 24 & Darling Downs & 43 & Central WA \\ 
		3 & Illawarra & 25 & Mackay \& Northern & 44 & Pilbara \\ 
		4 & Mid-North Coast & 26 & Yorke \& Lower North & 45 & Kimberley \\
		\cline{1-2}
		\multicolumn{2}{|c|}{\textbf{Country Victoria}} & 27 & Murray Lands & 46 & Mersey-Lyell \\
		\cline{1-2}
		6 & Barwon & 28 & South East & 47 & Northern Territory \\
		7 & Western District & 29 & South West & & \\
		8 & Central Highlands & 30 & Lower Southern WA & & \\
		9 & Loddon \& Goulbourn & 31 & Upper Southern WA & & \\
		10 & Gippsland & 32 & Midlands & & \\
		& & 33 & Northern Tasmania & & \\
		& & 34 & Darwin & & \\
		\hline
	\end{tabular}
\end{table}

To form an appropriate disaggregation structure for areas listed in Table~\ref{tab:1}, we collect seven capital cities and four broader regions into a group and name it by ``\textit{Region}". Thus, we have a three-level hierarchy, from top to bottom, consisting of the country total (Australia), 11 \textit{Region}s, and 47 \textit{Area}s, respectively. Together with sex in each area, there are in total 177 series across three levels of disaggregation listed in Table~\ref{tab:2}.

\begin{table}[!htbp]
	\tabcolsep 0.45in
	\centering
	\caption{Hierarchy of Australian mortality rates.}\label{tab:2}
	\begin{tabular}{@{}lr@{}}
		\toprule
		Level & Number of series \\
		\midrule
		Australia & 1 \\
		Sex & 2 \\
		Region & 11 \\
		Sex $\times$ Region & 22 \\
		Area & 47 \\
		Sex $\times$ Area & 94 \\
		\midrule
		All & 177 \\
		\bottomrule
	\end{tabular}
\end{table}

States and territories of Australia diversify significantly in area size, geographical features, and population. Most of Australia's population is concentrated in coastal (within 80 km from the coast wherever possible) regions, and the population within these regions is concentrated in capital cities in particular. In June 2010, around 14.3 million people, or approximately 64\% of Australia's population, lived in capital cities (including Darwin) \citep{ABS12}. Of the remainder population, over 6 million people (about one-third of the population) live in what referred to as regional and remote areas that are located with some distance away from major population centers and have extremely diverse geographical features \citep{AIHW03}. Mortality rates for populations living in the metropolitan, regional, and remote areas are expected to be influenced by different geography, climate, and environmental conditions. For instance, \cite{AIHW07} reported mortality rates in remote areas that have very little accessibility of goods, services, and opportunities for social interaction went up to 1.76 times in capital cities. Another reason for inconsistent death rates among metropolitan and non-metropolitan populations is that many of the occupations in regional and remote areas involve higher levels of risk than other occupations \citep{AIHW98}. For example, many regional and remote areas are primarily in the following high-risk industries: mining, transport, forestry, commercial fishing, and farming. 

\begin{figure}[!htb]
	\centering
	\includegraphics[width = \textwidth]{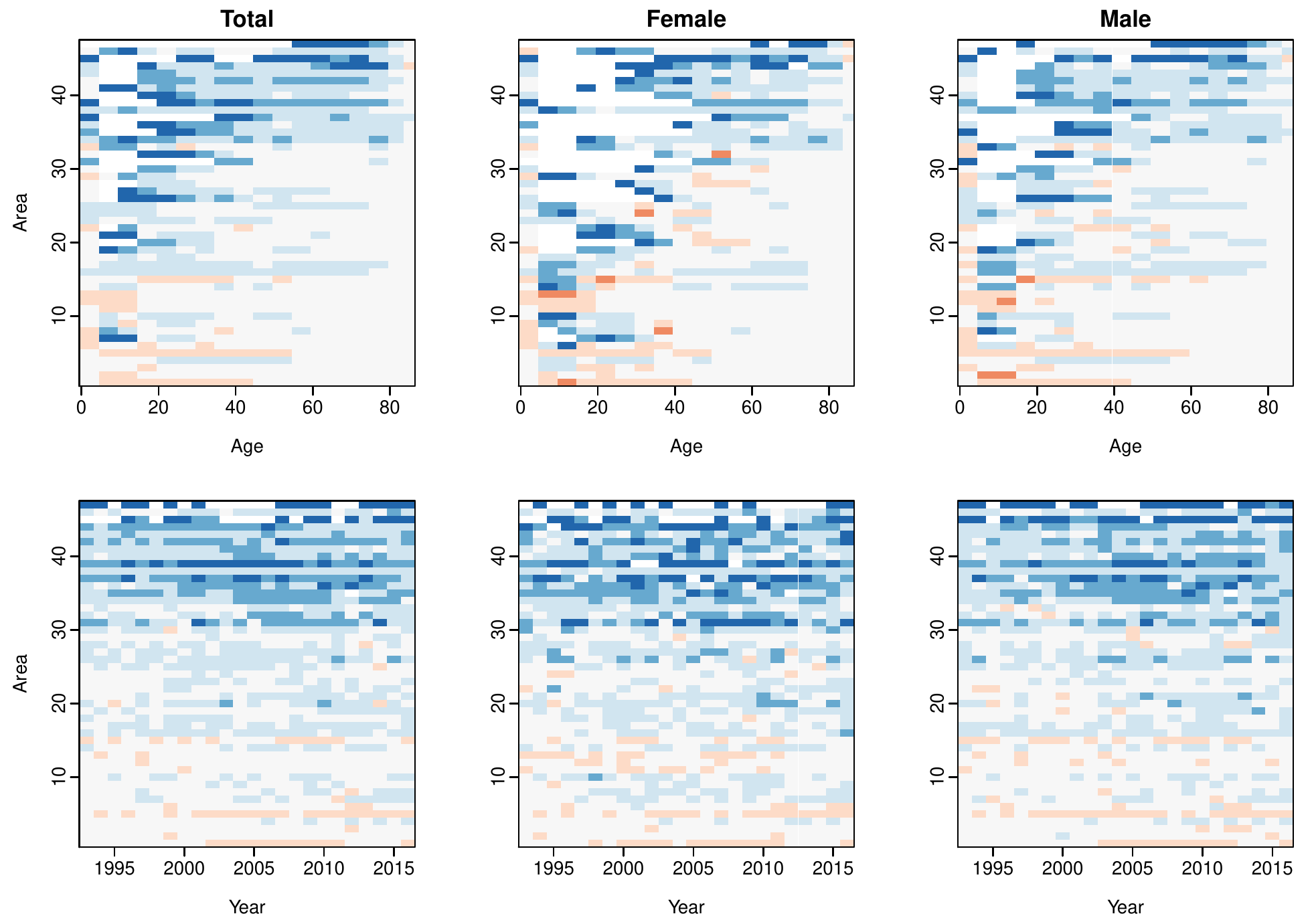}
	\caption{Image plots showing log ratios of mortality rates. The top panel shows mortality rates averaged over the years, while the bottom panel shows mortality rates averaged over the ages.}
	\label{fig: image}
\end{figure}

To visualize the differences between metropolitan and non-metropolitan mortality patterns, we divide the mortality rates of each area to death rates for Australia and then take the natural logarithm of the ratio. In Figure~\ref{fig: image}, positive log-ratios are represented in blue color, whereas negative values are represented in orange color, with the vertical axis indicating area numbers shown in Table~\ref{tab:1}. In the top row, the national mortality rates averaged over the period during 1993-2016 show that people living in capital cities have lower mortality rates than the country average, especially during the teenage years. In contrast, regional and remote areas, in general, have higher mortality rates for middle-to-old populations. The disadvantage of regional and remote areas regarding mortality is that people living in these areas have lower access to health services \citep{AIHW98}. In the bottom row, the mortality rates for each area and year, averaged over all ages, highlight a consistent trend that metropolitan populations enjoy lower mortality rates than non-metropolitan populations from 1993 to 2016. 

\begin{figure}[!htb]
 	\centering
 	\subfloat[Total series hierarchy]
 	{\includegraphics[width = 3in]{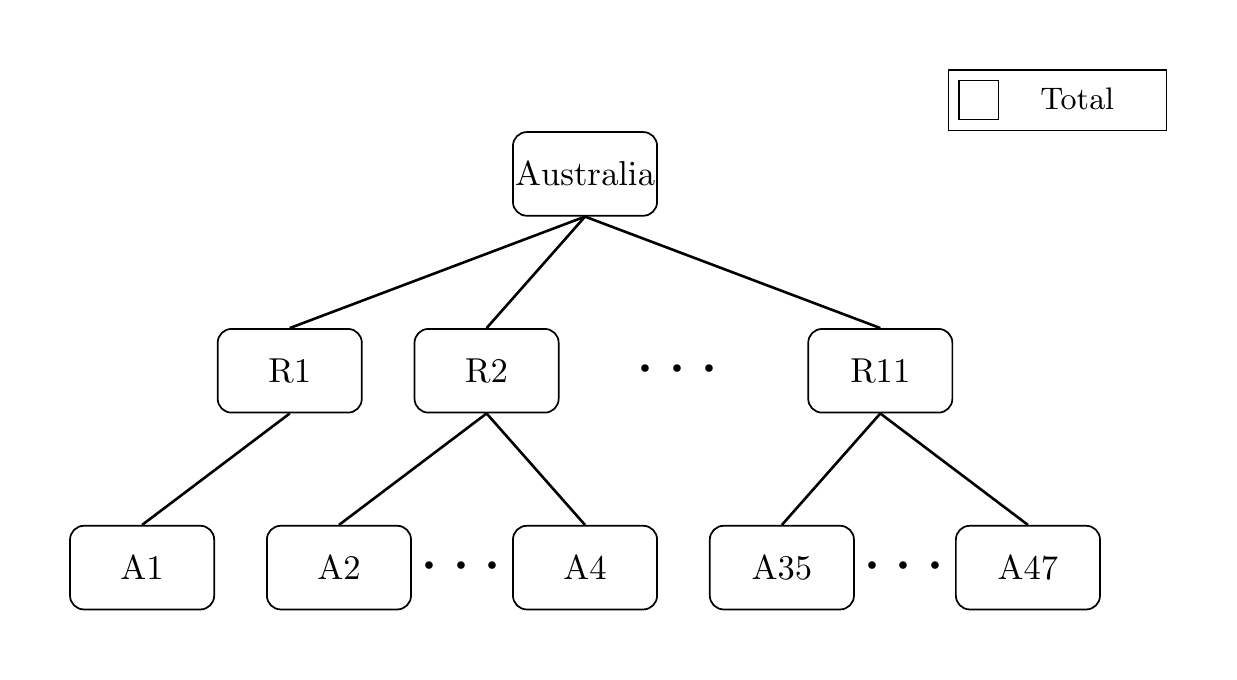}\label{fig:4a}} \qquad
 	\subfloat[Sex series hierarchy]
 	{\includegraphics[width = 3in]{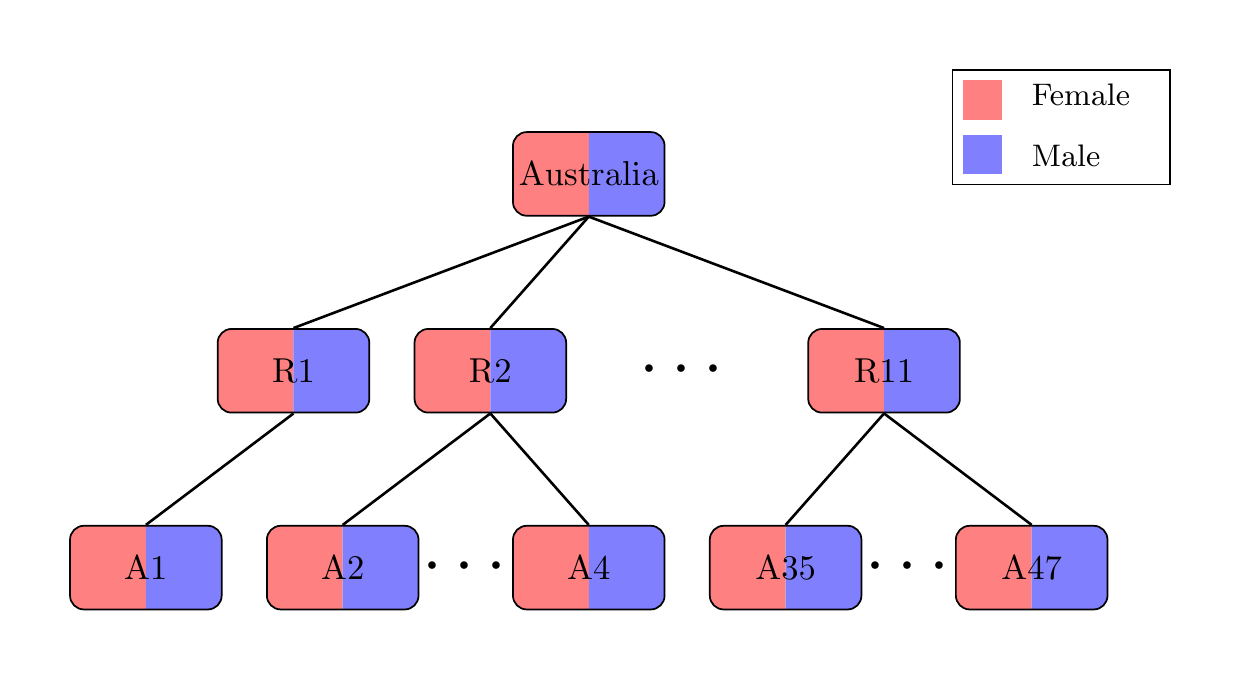}\label{fig:4b}} 
 	\caption{Hierarchy tree diagrams for the Australian mortality rates.}
 	\label{fig: structure}
 \end{figure}

Figure~\ref{fig: image} also shows that related populations tend to have similar mortality features. On the one hand, each of three categories, Capital City, Regional Australia, and Remote Australia, has specific mortality patterns explainable by living conditions of areas in that remoteness classification. On the other hand, females and males in each area across the country appear to show consistent mortality movements during 1993-2016. Considering the distinct mortality patterns of metropolitan and non-metropolitan populations, we jointly model total death rates of areas within each of 11 \textit{Region}s shown in Figure~\ref{fig:4a}, before collectively modeling total series for $R1, \cdots, R11$. Due to the close correlation of female and male populations in each Region and Area, we jointly consider sex series (colored by red and blue) in every node of Figure~\ref{fig:4b}. Tree diagrams illustrating possible disaggregation of Australian national and sub-national mortality rates are shown in Figure~\ref{fig: structure}. In Section~\ref{sec:method}, we will introduce the methodology of jointly modeling functional time series.

 \section{Methodology} \label{sec:method}

\subsection{Multivariate functional principal component analysis} \label{sec:3.2}

The sub-national mortality rates of Australia can be organized into sets of multivariate functional time series based on the combinations of geographical disaggregation factors and sex. Following \cite{CCY14}, we define  $\{ f^{(l)}(x) \}_{l = 1, \cdots, \omega }$ as a set of smoothed sub-national mortality functions, with $\omega$ representing the number of series under consideration. In our study, $\omega \in \{1, 2, 11, 22, 47, 94\}$ as shown in Table~\ref{tab:2}, and $\omega = 1$ corresponds to the special case of univariate functional time series. For $\omega \geq 2$, define $\bm{f}(x) = \left[f^{(1)}(x),\dots,f^{(\omega)}(x)\right]^{\top}$ as a vector in the Hilbert space. All elements of $\bm{f}(x)$ in the study are square-integrable functions defined on the same interval of age $\I = [0, 85]$.

Let $\mu^{(l)}(x) := \text{E}[f^{(l)}(x)]$ denote the mean function for the $l^{\text{th}}$ subpopulation. For $x, z\in \I$, the covariance function of $\bm{f}(x)$ has its $(l,j)^{\text{th}}$ elements defined as follows, where $1\leq l,j \leq \omega$:
 \begin{align*}
 C_{lj}(x, z) :=  \text{E} \Big\{ [f^{(l)}(x)-\mu^{(l)}(x)] [f^{(j)}(z) - \mu^{(j)}(z)] \Big\}  = \text{Cov}\left[f^{(l)}(x), f^{(j)}(z)\right].
 \end{align*}
Let $\bm{C}_l(x,z) = \left[C_{l1}(x,z), \cdots, C_{l \omega}(x,z) \right]^{\top} $ and $\bm{C}(x,z) = \{C_{lj}(x,z)\}$. We define an integral operator $\mathcal{A}: \mathbb{H} \rightarrow \mathbb{H} $ with the covariance kernel $\bm{C}(x,z)$, for any given $\bm{f} \in \mathbb{H}$, such that
 \begin{align*}
 	(\mathcal{A} \bm{f} ) (x) = \int \bm{C}(x,z) \bm{f} (z) dz = \begin{pmatrix}
 		\inp{\bm{C}_1(x,z)}{\bm{f}(z)} \\ \vdots \\  \inp{\bm{C}_{\omega}(x,z)}{\bm{f}(z)}   
 	\end{pmatrix},
 \end{align*}
where $\inp{\bm{C}_l(x,z)}{\bm{f}(z)} = \sum_{j=1}^{\omega}\inp{C_{lj}(x,z)}{f^{(j)}(z)} $. Since $\mathcal{A}$ is linear, an eigenvalue $\lambda$ and an eigenfunction $\bm{\phi}$ in $\mathbb{H}$ satisfy $(\mathcal{A} \bm{\phi} ) (x) = \lambda \bm{\phi}(x) $. Via Mercer's lemma, there exists orthonormal sequences $\{\bm{\phi}_k = [\phi_k^{(1)}, \cdots, \phi_k^{(\omega)}]^{\top} \}_{k=1,2,\cdots}$ of continuous functions in $\mathbb{H}$, and a non-increasing sequence $\lambda_k$ of positive numbers, such that
 \begin{align*}
 \bm{C}(x,z) = \sum_{k=1}^{\infty} \lambda_k \bm{\phi}_k(x)\bm{\phi}_k(z),
 \end{align*}
 with the $(l,j)^{\text{th}}$ element $C_{lj}(x,z)$ of $\bm{C}(x,z) $
 \begin{align*}
 	C_{lj}(x,z) = \sum_{k=1}^{\infty}\lambda_k \phi_{k}^{(l)}(x) \phi_{k}^{(j)}.
 \end{align*}
 By the separability of Hilbert spaces, the Karhunen-Lo\`{e}ve expansion of a stochastic process for the $l^{\text{th}}$ subpopulation can be expressed as
 \begin{align*}
 f^{(l)}(x) &= \mu^{(l)}(x) + \sum^{\infty}_{k=1}\beta^{(l)}_{k}\phi_k^{(l)}(x),
 \end{align*}
 where $\beta^{(l)}_{k}$ is the $k^{\text{th}}$ principal component score given by the projection of $f^{(l)}(x) - \mu^{(l)}(x)$ in the direction of eigenfunction $\phi_k^{(l)}$, that is, $\beta^{(l)}_{k} = \inp{f^{(l)}(x) - \mu^{(l)}(x)}{\phi_k^{(l)}(x)}$.

The proposed multivariate functional principal component analysis (MFPCA) as an extension of the widely used FPCA approach \citep[for theoretical, methodological, and applied aspects of FPCA, see, e.g.,][]{Hall11, Shang14, WCM16}, explicitly incorporate cross-covariance between subpopulation curves. We use this dimension reduction technique to effectively summarize the main features of each infinite-dimensional curves considered by its finite key elements, and form a base of functional principal component regression. 

\subsection{Forecasting via functional principal component regression} \label{sec:3.3}

A time series of smoothed functions corresponding to multiple subpopulations $\{\bm{f}_1(x),\cdots, \bm{f}_n(x)\}$ can be decomposed into orthogonal functional principal components and their associated scores as
 \begin{align}
 	f_t^{(l)}(x) &= \mu^{(l)}(x) + \sum^{\infty}_{k=1}\beta^{(l)}_{t,k}\phi_k^{(l)}(x) \notag \\
 	&= \mu^{(l)}(x) + \sum_{k=1}^{K}\beta^{(l)}_{t,k}\phi^{(l)}_k(x) + e^{(l)}_t(x), \label{eq:FPCA}
 \end{align}
 where $\mu^{(l)}(x)$ is the mean function for the $l^{\text{th}}$ subpopulation; $\Big\{\phi_1^{(l)}(x),\dots,\phi_{K}^{(l)}(x)\Big\}$ is a set of the first $K$ functional principal components; $\Big\{\bm{\beta}_1^{(l)},\cdots,\bm{\beta}^{(l)}_{K}\Big\}$, with elements $\bm{\beta}_1^{(l)} = \left[\beta_{1,1}^{(l)},\dots,\beta_{n,1}^{(l)}\right]^{\top}$, denotes a set of principal component scores for the $l^{\text{th}}$ subpopulation; $e_t^{(l)}(x)$ denotes the model truncation error function with mean zero and finite variance for the $l^{\text{th}}$ subpopulation;  and $K<n$ is the number of retained principal components. 

Following \cite{SH16}, the number of retained components is determined as the minimum that reaches 95\% of total variance explained by the leading components\footnote{Several alternative methods are used for selecting the number of functional principal components $K$, such as those of \cite{Chiou12, YMW05, RS91, HV06}.}, such that
\begin{equation*}
 K = \argmin_{K: K\geq 1}\left\{\sum^K_{k=1}\widehat{\lambda}_k\Bigg/\sum^{\infty}_{k=1}\widehat{\lambda}_k\mathds{1}_{\left\{\widehat{\lambda}_k>0\right\}}\geq 0.95\right\},
\end{equation*}
where $\mathds{1}\{\cdot\}$ represents the binary indicator function. Since all individual elements of $\bm{C}(x,z)$ share the same set of eigenvalues, only one common $K$ value is needed in dimension-reduction for all subpopulations.

Expansion~\eqref{eq:FPCA} facilitates dimension reduction since the first $K$ terms often provide a good approximation to the infinite sums, and thus the information contained in time series $\{f_1^{(l)}(x),\cdots,f_n^{(l)}(x) \}$ can be adequately summarized by the $K$-dimensional vector $\left\{\bm{\beta}^{(l)}_1,\dots,\bm{\beta}^{(l)}_{K}\right\}$. When collectively considering all $\omega$ time series, truncating at the first $K^{\text{th}}$ functional principal components yields the approximation in its matrix formulation as
 \begin{equation}
 \bm{f}_t(x) = \bm{\Phi}(x)\bm{\beta}_t^{\top},
 \label{eq:ft}
 \end{equation}
 where $\bm{\beta}_t = \left\{ \beta_{t,1}^{(1)}, \dots, \beta_{t,K}^{(1)}, \beta_{t,1}^{(2)},\dots,\beta_{t,K}^{(2)},\dots, \beta_{t,1}^{(\omega)},\dots,\beta_{t,K}^{(\omega)}\right\}$ is the vector of the basis expansion coefficients, and 
 \begin{equation*}
 \bm{\Phi}(x) = \left( \begin{array}{ccccccccc}
 \phi_1^{(1)}(x) & \cdots & \phi_{K}^{(1)}(x) & 0 & \cdots & 0 & 0 & \cdots & 0 \\
 0 & \cdots & 0 & \phi_1^{(2)}(x) & \cdots & \phi_{K}^{(2)}(x) & 0 & \cdots & 0 \\
 \vdots & \vdots & \vdots & \vdots & \vdots & \vdots & \vdots & \vdots & \vdots \\
  0 & \cdots & 0 & 0 & \cdots & 0 & \phi_1^{(\omega)}(x) & \cdots & \phi_{K}^{(\omega)}(x)  \end{array} \right)_{\omega\times (K\times \omega)}.
 \end{equation*}

For dense and regularly spaced functional time series, such as age-specific mortality rates,  the mean function $\widehat{\bm{\mu}}(x) = \frac{1}{n}\sum_{t=1}^{n}\bm{f}_t(x) $ and the  covariance function $\widehat{\bm{C}}(x,z)$ can be empirically estimated. Based on Eq.~\eqref{eq:ft}, we extract empirical basis functions from the empirical covariance function $\widehat{\bm{\Phi}}(x)$, and then make $h$-step-ahead point forecasts as
 \begin{align*}
 \widehat{\bm{f}}_{n+h|n} & = \text{E}[\bm{f}_{n+h}(x)|\bm{f}_1(x),\cdots,\bm{f}_n(x);\widehat{\bm{\Phi}}(x)] \\
 & = \widehat{\bm{\mu}}(x) + \sum_{k=1}^{K} \widehat{\bm{\beta}}_{n+h|n,k} \widehat{\bm{\Phi}}(x),
 \end{align*}
 where the forecast principal component score $\widehat{\bm{\beta}}_{n+h|n,k}$ is obtained using the automated autoregressive integrated moving average (ARIMA) model-fitting algorithm of \cite{HK08}. More details about forecasting functional principal component scores can be found in \cite{HS09}.
 
 We adopt a univariate time series forecasting method of \cite{HS09} to obtain the forecast principal component score $\widehat{\beta}_{n+h|n,k}$. This univariate time series forecasting method can model a time series with a stochastic trend component. Since the yearly age-specific mortality rates do not contain seasonality, the ARIMA has the general form of
  \begin{equation*}
 (1-\tau_1 B - \cdots - \tau_p B^p)(1-B)^d \bm{\beta}_k = \varphi + (1+\nu_1 B + \cdots + \nu_q B^q)\bm{w}_k,
 \end{equation*}
  where $\varphi$ represents the intercept, $(\tau_1,\dots,\tau_p)$ denote the coefficients associated with the autoregressive component, $\bm{\beta}_k = \left\lbrace \beta_{1,k}, \cdots, \beta_{n,k} \right\rbrace $ represents principal component scores, $(\nu_1,\cdots,\nu_q)$ denote the coefficients associated with the moving average component, $B$ denotes the backshift operator, $d$ denotes the differencing operator, and $\bm{w}_k = \{w_{1,k},\dots,w_{n,k}\}$ represents a white-noise error term. We use the automatic algorithm of \cite{HK08} to choose the optimal autoregressive order $p$, moving average order $q$, and difference order $d$. The value of $d$ is selected based on successive Kwiatkowski-Phillips-Schmidt-Shin (KPSS) unit root tests \citep{KPSS92}. KPSS tests are used for testing the null hypothesis that an observable time series is stationary around a deterministic trend. We first test the original time series for a unit root; if the test result is significant, then we test the differenced time series for a unit root. The procedure continues until we obtain our first insignificant result. Once the order $d$ is determined, the orders of $p$ and $q$ are selected based on the optimal AIC with a correction for a small finite sample size \citep{Akaike74}. Having identified the optimal ARIMA model, the maximum likelihood method can be used to estimate the parameters.

\section{Forecast reconciliation for grouped functional time series} \label{sec:recon}

\subsection{Notation} \label{sec:4.1}

For both hierarchies illustrated in Figure~\ref{fig: structure}, we denote a particular disaggregated series using the notation $\text{G} \ast \text{S}$, meaning the geographical area G and the sex S. For instance, $\text{R}_1 \ast \text{F}$ denotes females in Region 1, $\text{A}_1 \ast \text{T}$ denotes females and males in Area 1, and $\text{Australia} \ast \text{M}$ denotes males in Australia. Let $E_{\text{G}\ast \text{S}, t}(x)$ denote the exposure to risk for series $\text{G}\ast \text{S}$ in year $t$ and age $x$, and let $D_{\text{G}\ast \text{S}, t}(x)$ be the number of deaths for series $\text{G}\ast \text{S}$ in year $t$ and age $x$. Then, age-specific mortality rate is given by $R_{\text{G}\ast\text{S}, t}(x) = D_{\text{G}\ast \text{S}, t}(x)/E_{\text{G}\ast \text{S}, t}(x)$. Dropping the age variable $(x)$ allows us to express the national and sub-national mortality series in a matrix multiplication as

\begin{small}
 \arraycolsep=0.05cm
 \[
 \underbrace{ \left[
 \begin{array}{l}
 R_{\text{Australia}\ast \text{T},t} \\
 R_{\textcolor{red}{\text{Australia}\ast \text{F},t}} \\
 R_{\textcolor{red}{\text{Australia}\ast \text{M},t}} \\
 R_{\textcolor{a0}{\text{R1}\ast \text{T},t}} \\
 R_{\textcolor{a0}{\text{R2}\ast \text{T},t}} \\
 \vdots \\
 R_{\textcolor{a0}{\text{R11}\ast \text{T},t}} \\
 R_{\textcolor{blue-violet}{\text{R1}\ast \text{F},t}} \\
 R_{\textcolor{blue-violet}{\text{R2}\ast \text{F},t}} \\
 \vdots \\
 R_{\textcolor{blue-violet}{\text{R11}\ast \text{F},t}} \\
 R_{\textcolor{burntorange}{\text{R1}\ast \text{M},t}} \\
 R_{\textcolor{burntorange}{\text{R2}\ast \text{M},t}} \\
 \vdots \\
 R_{\textcolor{burntorange}{\text{R11}\ast \text{M},t}} \\
 R_{\textcolor{blue}{\text{A1}\ast \text{T},t}} \\
 R_{\textcolor{blue}{\text{A2}\ast \text{T},t}} \\
 \vdots \\
 R_{\textcolor{blue}{\text{A47}\ast \text{T},t}} \\
 R_{\textcolor{purple}{\text{A1}\ast \text{F},t}} \\
 R_{\textcolor{purple}{\text{A1}\ast \text{M},t}} \\
 R_{\textcolor{purple}{\text{A2}\ast \text{F},t}} \\
 R_{\textcolor{purple}{\text{A2}\ast \text{M},t}} \\
 \vdots \\
 R_{\textcolor{purple}{\text{A47}\ast \text{F},t}} \\
 R_{\textcolor{purple}{\text{A47}\ast \text{M},t}} \\ \end{array}
 \right]}_{\bm{R}_t} =
 \underbrace{\left[
 \begin{array}{ccccccccccc}
 \frac{E_{\text{A1}\ast \text{F},t}}{E_{\text{Australia}\ast \text{T},t}} & \frac{E_{\text{A1}\ast \text{M},t}}{E_{\text{Australia}\ast \text{T},t}} & \frac{E_{\text{A2}\ast \text{F},t}}{E_{\text{Australia}\ast \text{T},t}} & \frac{E_{\text{A2}\ast \text{M},t}}{E_{\text{Australia}\ast \text{T},t}}  & \frac{E_{\text{A3}\ast \text{F},t}}{E_{\text{Australia}\ast \text{T},t}} & \frac{E_{\text{A3}\ast \text{M},t}}{E_{\text{Australia}\ast \text{T},t}} & \cdots & \frac{E_{\text{A47}\ast \text{F},t}}{E_{\text{Australia}\ast \text{T},t}} & \frac{E_{\text{A47}\ast \text{M},t}}{E_{\text{Australia}\ast \text{T},t}} \\
 \textcolor{red}{\frac{E_{\text{A1}\ast \text{F},t}}{E_{\text{Australia}\ast \text{F},t}}} & \textcolor{red}{0} & \textcolor{red}{\frac{E_{\text{A2}\ast \text{F},t}}{E_{\text{Australia}\ast \text{F},t}}} & \textcolor{red}{0} & \textcolor{red}{\frac{E_{\text{A3}\ast \text{F},t}}{E_{\text{Australia}\ast \text{F},t}}} & \textcolor{red}{0} & \cdots & \textcolor{red}{\frac{E_{\text{A47}\ast \text{F},t}}{E_{\text{Australia}\ast \text{F},t}}} & \textcolor{red}{0} \\
 \textcolor{red}{0} & \textcolor{red}{\frac{E_{\text{A1}\ast \text{M},t}}{E_{\text{Australia}\ast \text{M},t}}}  & \textcolor{red}{0} & \textcolor{red}{\frac{E_{\text{A2}\ast \text{M},t}}{E_{\text{Australia}\ast \text{M},t}}} & \textcolor{red}{0} & \textcolor{red}{\frac{E_{\text{A3}\ast \text{M},t}}{E_{\text{Australia}\ast \text{M},t}}} & \cdots & \textcolor{red}{0} & \textcolor{red}{\frac{E_{\text{A47}\ast \text{M},t}}{E_{\text{Australia}\ast \text{M},t}}} \\
 \textcolor{a0}{\frac{E_{\text{A1}\ast \text{F},t}}{E_{\text{R1}\ast \text{T},t}}} & \textcolor{a0}{\frac{E_{\text{A1}\ast \text{M},t}}{E_{\text{R1}\ast \text{T},t}}} & \textcolor{a0}{0} & \textcolor{a0}{0} & \textcolor{a0}{0} & \textcolor{a0}{0} & \cdots  & \textcolor{a0}{0} & \textcolor{a0}{0} \\
 \textcolor{a0}{0} & \textcolor{a0}{0} & \textcolor{a0}{\frac{E_{\text{A2}\ast \text{F},t}}{E_{\text{R2} \ast \text{T},t}}} & \textcolor{a0}{\frac{E_{\text{A2}\ast \text{M},t}}{E_{\text{R2}\ast \text{T},t}}} & \textcolor{a0}{\frac{E_{\text{A3}\ast \text{F},t}}{E_{\text{R2}\ast \text{T},t}}} & \textcolor{a0}{\frac{E_{\text{A3}\ast \text{M},t}}{E_{\text{R2}\ast \text{T},t}}} & \cdots & \textcolor{a0}{0} & \textcolor{a0}{0} \\
 \vdots & \vdots & \vdots & \vdots & \vdots & \vdots & \cdots & \vdots & \vdots \\
 \textcolor{a0}{0} & \textcolor{a0}{0} & \textcolor{a0}{0} & \textcolor{a0}{0} & \textcolor{a0}{0} & \textcolor{a0}{0} & \cdots & \textcolor{a0}{\frac{E_{\text{A47}\ast \text{F},t}}{E_{\text{R8}\ast \text{T},t}}} & \textcolor{a0}{\frac{E_{\text{A47}\ast \text{M},t}}{E_{\text{R8}\ast \text{T},t}}} \\
 \textcolor{blue-violet}{\frac{E_{\text{A1}\ast \text{F},t}}{E_{\text{R1}\ast \text{F},t}}} & \textcolor{blue-violet}{0} & \textcolor{blue-violet}{0} & \textcolor{blue-violet}{0} & \textcolor{blue-violet}{0} & \textcolor{blue-violet}{0} &  \cdots & \textcolor{blue-violet}{0} & \textcolor{blue-violet}{0} \\
 \textcolor{blue-violet}{0} & \textcolor{blue-violet}{0} & \textcolor{blue-violet}{\frac{E_{\text{A2}\ast \text{F},t}}{E_{\text{R2}\ast \text{F},t}}} & \textcolor{blue-violet}{0} & \textcolor{blue-violet}{\frac{E_{\text{A3}\ast \text{F},t}}{E_{\text{R2}\ast \text{F},t}}} & \textcolor{blue-violet}{0} & \cdots & \textcolor{blue-violet}{0} & \textcolor{blue-violet}{0}  \\
 \vdots & \vdots & \vdots & \vdots & \vdots & \vdots & \cdots & \vdots & \vdots \\
 \textcolor{blue-violet}{0} & \textcolor{blue-violet}{0}  & \textcolor{blue-violet}{0}  & \textcolor{blue-violet}{0}  & \textcolor{blue-violet}{0}  & \textcolor{blue-violet}{0}  & \cdots & \textcolor{blue-violet}{\frac{E_{\text{A47}\ast \text{F},t}}{E_{\text{R8}\ast \text{F},t}}} & \textcolor{blue-violet}{0}\\
 \textcolor{burntorange}{0} & \textcolor{burntorange}{\frac{E_{\text{A1}\ast \text{M},t}}{E_{\text{R1}\ast \text{M},t}}} & \textcolor{burntorange}{0} &\textcolor{burntorange}{0} & \textcolor{burntorange}{0} & \textcolor{burntorange}{0} & \cdots & \textcolor{burntorange}{0} & \textcolor{burntorange}{0} \\
 \textcolor{burntorange}{0} & \textcolor{burntorange}{0} & \textcolor{burntorange}{0} & \textcolor{burntorange}{\frac{E_{\text{A2}\ast \text{M},t}}{E_{\text{R2}\ast \text{M},t}}} & \textcolor{burntorange}{0} & \textcolor{burntorange}{\frac{E_{\text{A3}\ast \text{M},t}}{E_{\text{R2}\ast \text{M},t}}} & \cdots & \textcolor{burntorange}{0} & \textcolor{burntorange}{0} \\
 \vdots & \vdots & \vdots & \vdots & \vdots & \vdots & \cdots & \vdots & \vdots \\
 \textcolor{burntorange}{0} & \textcolor{burntorange}{0} & \textcolor{burntorange}{0} & \textcolor{burntorange}{0} & \textcolor{burntorange}{0} & \textcolor{burntorange}{0} & \cdots & \textcolor{burntorange}{0} & \textcolor{burntorange}{\frac{E_{\text{A47}\ast \text{M},t}}{E_{\text{R8}\ast \text{M},t}}} \\
 \textcolor{blue}{\frac{E_{\text{A1}\ast \text{F},t}}{E_{\text{A1}\ast \text{T},t}}} & \textcolor{blue}{\frac{E_{\text{A1}\ast \text{M},t}}{E_{\text{A1}\ast \text{T},t}}} & \textcolor{blue}{0} & \textcolor{blue}{0} & \textcolor{blue}{0} & \textcolor{blue}{0} & \cdots & \textcolor{blue}{0} & \textcolor{blue}{0} \\
 \textcolor{blue}{0} & \textcolor{blue}{0}  &  \textcolor{blue}{\frac{E_{\text{A2}\ast \text{F},t}}{E_{\text{A2}\ast \text{T},t}}} & \textcolor{blue}{\frac{E_{\text{A2}\ast \text{M},t}}{E_{\text{A2}\ast \text{T},t}}} & \textcolor{blue}{0} & \textcolor{blue}{0}  & \cdots & \textcolor{blue}{0} & \textcolor{blue}{0} \\
 \vdots & \vdots & \vdots & \vdots & \vdots & \vdots & \cdots & \vdots & \vdots \\
 \textcolor{blue}{0} & \textcolor{blue}{0} & \textcolor{blue}{0} & \textcolor{blue}{0} & \textcolor{blue}{0} & \textcolor{blue}{0} & \cdots & \textcolor{blue}{\frac{E_{\text{A47}\ast \text{F},t}}{E_{\text{A47}\ast \text{T},t}}} & \textcolor{blue}{\frac{E_{\text{A47}\ast \text{M},t}}{E_{\text{A47}\ast \text{T},t}}} \\
 \textcolor{purple}{1} & \textcolor{purple}{0} & \textcolor{purple}{0} & \textcolor{purple}{0} & \textcolor{purple}{0} & \textcolor{purple}{0} & \cdots & \textcolor{purple}{0} & \textcolor{purple}{0} \\
 \textcolor{purple}{0} & \textcolor{purple}{1} & \textcolor{purple}{0} & \textcolor{purple}{0} & \textcolor{purple}{0} & \textcolor{purple}{0} & \cdots & \textcolor{purple}{0} & \textcolor{purple}{0} \\
 \textcolor{purple}{0} & \textcolor{purple}{0} & \textcolor{purple}{1} & \textcolor{purple}{0} & \textcolor{purple}{0} & \textcolor{purple}{0} & \cdots & \textcolor{purple}{0} & \textcolor{purple}{0} \\
 \textcolor{purple}{0} & \textcolor{purple}{0} & \textcolor{purple}{0} & \textcolor{purple}{1} & \textcolor{purple}{0} & \textcolor{purple}{0} & \cdots & \textcolor{purple}{0} & \textcolor{purple}{0} \\
 \vdots & \vdots & \vdots & \vdots & \vdots & \vdots & \cdots & \vdots & \vdots  \\
 \textcolor{purple}{0} & \textcolor{purple}{0} & \textcolor{purple}{0} & \textcolor{purple}{0} & \textcolor{purple}{0} & \textcolor{purple}{0} & \cdots  & \textcolor{purple}{1} & \textcolor{purple}{0}\\
 \textcolor{purple}{0} & \textcolor{purple}{0} & \textcolor{purple}{0} & \textcolor{purple}{0} & \textcolor{purple}{0} & \textcolor{purple}{0} & \cdots & \textcolor{purple}{0} & \textcolor{purple}{1} \\
 \end{array}
 \right]}_{\bm{S}_t}
 \underbrace{\left[
 \begin{array}{l}
 R_{\text{A1}\ast \text{F},t} \\
 R_{\text{A1}\ast \text{M},t} \\
 R_{\text{A2}\ast \text{F},t} \\
 R_{\text{A2}\ast \text{M},t} \\
 \vdots \\
 R_{\text{A47}\ast \text{F},t} \\
 R_{\text{A47}\ast \text{M},t} \\
 \end{array}
 \right]}_{\bm{b}_t}
 \]
 \end{small}
\hspace{-.05in} or $\bm{R}_t = \bm{S}_t\bm{b}_t$, where $\bm{R}_t$ is a vector containing all series at all levels of disaggregation, $\bm{b}_t$ is a vector of the most disaggregated series, and $\bm{S}_t$ shows how the two are related. We present a brief review of three methods for forecast reconciliation based on this equation in Sections~\ref{sec:4.2} to~\ref{sec:4.5}.

\subsection{The bottom-up (BU) method} \label{sec:4.2}

A widely used approach to reconcile grouped functional time series is the bottom-up method \citep[see, e.g.,][]{DM92, ZT00}. This method has the agreeable feature that it is intuitive and straightforward and always results in forecasts that satisfy the same group structure as the original data. 

First, we adopt the multivariate functional mortality forecasting method to obtain $h$-step-ahead base forecasts for the most disaggregated series, denoting it by $\widehat{\bm{b}}_{n+h}$ = $\Big[\widehat{R}_{\text{A}_1\ast \text{F}, n+h},$ $\widehat{R}_{\text{A}_1\ast \text{M}, n+h}$, $\widehat{R}_{\text{A}_2\ast \text{F}, n+h}$, $\widehat{R}_{\text{A}_2\ast \text{M}, n+h}$, $\cdots,$ $\widehat{R}_{\text{A}_{47}\ast \text{F}, n+h}, \widehat{R}_{\text{A}_{47}\ast \text{M}, n+h}\Big]^{\top}$. Then, we use empirical ratios to form the $\bm{S}_t$ for all $t \in [1993, 2016]$  \citep[see also][]{SH16b}. Thus, we obtain reconciled forecasts for all series
\begin{equation*}
\overline{\bm{R}}_{n+h} = \bm{S}_{n+h}\widehat{\bm{b}}_{n+h},
\end{equation*}
where $\overline{\bm{R}}_{n+h}$ denotes the reconciled forecasts.

\subsection{The optimal combination (OP) method} \label{sec:4.3}

Instead of considering only the bottom-level series, \cite{HAA+11} proposed a method in which base forecasts for all aggregated and disaggregated series are computed successively. Then the resulting forecasts are combined through linear regression. The reconciled forecasts are as close as possible to the base forecasts and aggregate consistently within the group.

The method is derived by writing the base forecasts as the response variable of the linear regression:
\begin{equation*}
\widehat{\bm{R}}_{n+h} = \bm{S}_{n+h}\bm{\beta}_{n+h}+\bm{\epsilon}_{n+h},
\end{equation*}
where $\widehat{\bm{R}}_{n+h}$ is a matrix of $h$-step-ahead base forecasts for all series, stacked in the same order as for the original data; $\bm{\beta}_{n+h} = \text{E}[\bm{b}_{n+h}|\bm{R}_1,\dots,\bm{R}_n]$ is the unknown mean of the forecast distributions of the most disaggregated series; and $\bm{\epsilon}_{n+h}$ represents the reconciliation error which is independent of past observations, with mean zero and variance-covariance matrix $\bm{\Sigma}_h := \text{var}(\bm{\epsilon}_{n+h})$. 

We follow \cite{HAA+11} and \cite{Hyndman2016-wp} and estimate the regression coefficient in a weighted least squares fashion:
\begin{equation*}
\widehat{\bm{\beta}}_{n+h} = \left(\bm{S}_{n+h}^{\top}\bm{W}^{-1}\bm{S}_{n+h}\right)^{-1}\bm{S}_{n+h}^{\top}\bm{W}^{-1}\widehat{\bm{R}}_{n+h},
\end{equation*}
where $\bm{W}$ is a weight matrix. To guarantee the invertibility of $\bm{W}$, we assume $\bm{W} = c \times \bm{I}$, where $c$ is any constant and $\bm{I}$ denotes the identity matrix. Thus, the reconciled forecasts can be obtained via the ordinary least-squares method as
\begin{align*}
\overline{\bm{R}}_{n+h} = \bm{S}_{n+h}\widehat{\bm{\beta}}_{n+h} &= \bm{S}_{n+h}\left[\bm{S}_{n+h}^{\top}(c\times \bm{I})^{-1}\bm{S}_{n+h}\right]^{-1}\bm{S}_{n+h}^{\top}(c\times \bm{I})^{-1}\widehat{\bm{R}}_{n+h} \\
&= \bm{S}_{n+h}\left(\bm{S}_{n+h}^{\top}\bm{S}_{n+h}\right)^{-1}\bm{S}_{n+h}^{\top}\widehat{\bm{R}}_{n+h}.
\end{align*}
By construction, these are aggregate consistent and involve a combination of all the base forecasts. They are also unbiased since $\text{E}[\overline{\bm{R}}_{n+h}] = \bm{S}_{n+h}\bm{\beta}_{n+h}$. 

\subsection{The trace minimization (MinT) method} \label{sec:4.4}

The optimal combination method attempts only to minimize the variance of reconciliation errors in linear regression. \cite{WAH19} proposed an improved approach of finding coherent forecasts for the grouped functional time series. By minimizing the sum of variances of reconciliation errors, essentially minimizing trace of the variance matrix, the new approach produces coherent forecasts across the entire collection of time series. Since the variance-covariance matrix of out-of-sample reconciliation errors is often not known and not identifiable in practice, the MinT method attempts to approximate $\bm{\Sigma}_h$ within-sample base forecast errors.

Define the $h$-step-ahead in-sample error of base forecasts as 
\begin{align*}
\widehat{\bm{e}}_{t+h} = \bm{R}_{t+h} - \widehat{\bm{R}}_{t+h},
\end{align*}
and the corresponding error of reconciled forecasts using information up to and including time $t$ as 
 \begin{align*}
 	\widetilde{\bm{e}}_{t+h} = \bm{R}_{t+h} - \overline{\bm{R}}_{t+h},
 \end{align*}
 with $t = 1, \cdots, n-h$. \cite{WAH19} proves that the variance-covariance matrix of $\widetilde{\bm{e}}_{t+h}$ can be written as
 \begin{align*}
 	\text{var}[ \bm{R}_{t+h} - \overline{\bm{R}}_{t+h}|\bm{R}_1, \cdots, \bm{R}_t] = \bm{S}_{t+h}\bm{P}\bm{W}_h\bm{S}_{t+h}^{\top}\bm{P}^{\top},
 \end{align*}
 where $\bm{P}$ is a projection matrix satisfying $\bm{S}_{t+h}\bm{P} \bm{S}_{t+h} = \bm{S}_{t+h}$, and $\bm{W}_h = \text{E}[\widehat{\bm{e}}_{t+h} \widehat{\bm{e}}_{t+h}^{\top}|\bm{R}_1, \cdots, \bm{R}_t]$ is the variance-covariance matrix of the $h$-step-ahead base forecast errors. By minimizing the trace of $\bm{S}_{t+h}\bm{P}\bm{W}_h\bm{S}_{t+h}^{\top}\bm{P}^{\top}$, we obtain
 \begin{align*}
 	\bm{P} = \left(\bm{S}_{t+h}^{\top}{\bm{W}}_h^{-1}\bm{S}_{t+h}\right)^{-1}\bm{S}_{t+h}^{\top}{\bm{W}}_h^{-1},
 \end{align*}
 leading to the reconciled forecasts from the MinT approach:
 \begin{align*}
 	\overline{\bm{R}}_{n+h}  = \bm{S}_{t+h}\left(\bm{S}_{t+h}^{\top}{\bm{W}}_h^{-1}\bm{S}_{t+h}\right)^{-1}\bm{S}_{t+h}^{\top}{\bm{W}}_h^{-1} \widehat{\bm{R}}_{n+h}.
 \end{align*}
 In this equation, $\bm{W}_h$ is the covariance matrix of the base forecast errors. Our study is estimated by shirking off-diagonal entries of the unbiased sample covariance estimator of the in-sample 1-step-ahead base forecast errors towards targets on its diagonal; for more details together with alternative estimations of $\bm{W}_h$, consult Section 2.4 of \cite{WAH19}.
 
\subsection{The forecast combination (Comb\_av) method} \label{sec:4.5}

It is possible to combine the forecast mentioned above reconciliation methods to reduce bias, variance, and uncertainty of forecasts. \cite{Shang2020} recently considered a forecast combination method that computes the averaged forecast for horizon $h$ as 
\begin{align*}
\overline{\bm{R}}_{n+h}^{\text{comb}} = \sum_{g=1}^{G}w_g\overline{\bm{R}}_{n+h}^{\text{g}},
\end{align*}
where $\overline{\bm{R}}_{n+h}^{\text{g}}$ denotes forecasts obtained via a reconciliation method, and $\left\lbrace w_1, \cdots, w_g \right\rbrace $ are weights summing to 1. Following \cite{Shang2020}, we adopt a simple average combination method that applies equal weighting on the BU, OP and MinT reconciled forecasts as
\begin{align*}
\overline{\bm{R}}_{n+h}^{\text{Comb\_av}} = \frac{1}{3}[\overline{\bm{R}}_{n+h}^{\text{BU}} + \overline{\bm{R}}_{n+h}^{\text{OP}} + \overline{\bm{R}}_{n+h}^{\text{MinT}}].
\end{align*}

\section{Empirical application results} \label{sec:result}

The Multivariate functional time series forecasting method, and a univariate method of \cite{SH16}, are applied to Australian age-specific mortality rates to obtain base forecasts following the hierarchy structures in Figure~\ref{fig: structure}. We then conduct reconciliation via the BU, OP, MinT, and Comb\_av methods. To assess model and parameter stabilities over time, we consider an expanding window analysis of considered time series models \citep[see][Chapter 9 for details]{ZW06}. Specifically, we initially use the observed mortality curves from 1993 to 2011 to produce one- to 5-step-ahead point forecasts. Through an expanding window approach, we re-estimate the parameters of considered models using the first 20 years of observations from 1993 to 2002 and make one- to 4-step-ahead forecasts using re-estimated models. The process is iterated with the sample size increased by one year until the end of the data period in 2016. This process produces five one-step-ahead forecasts, four two-step-ahead forecasts, and so on, up to one five-step-ahead forecast. We evaluate the point forecast accuracy and report our study results.

\subsection{Point forecast error measures} \label{sec:5.1}

To evaluate the point forecast accuracy, we compute the mean absolute forecast error (MAFE), and the root mean squared forecast error (RMSFE). Because the raw Australian sub-national mortality rates cover 18 age groups and contain missing values for some areas, we define the MAFE for each series $j$ as 
\begin{align*}
\text{MAFE}_j(h) &= \frac{1}{18\times (6-h)}\sum^{5}_{\varsigma = h}\sum^{18}_{i=1}\left| \left[f^{(j)}_{n+\varsigma}(x_i) - \widehat{f}^{(j)}_{n+\varsigma}(x_i)\right]\mathds{1}_{\left\{f^{(j)}_{n+\varsigma}(x_i)>0\right\}}\right| ,
\end{align*}
where $f^{(j)}_{n+\varsigma}(x_i)$ represents the actual holdout sample for the $i$\textsuperscript{th} age and $\varsigma$\textsuperscript{th} curve of the forecasting period, and $\widehat{f}^{(j)}_{n+\varsigma}(x_i)$ is the corresponding point forecasts; $\mathds{1}_{\left\{f^{(j)}_{n+\varsigma}(x_i)>0\right\}}$ is to exclude missing death rate observations with $\mathds{1}_{\left\lbrace \cdot \right\rbrace }$ representing the binary indicator function. For each series $j$, the RMSFE is given by
\begin{align*}
\text{RMSFE}_j(h) &= \sqrt{\frac{1}{18\times (6-h)}\sum^{5}_{\varsigma = h}\sum^{18}_{i=1} \left[f^{(j)}_{n+\varsigma}(x_i) - \widehat{f}^{(j)}_{n+\varsigma}(x_i)\right]^2\mathds{1}_{\left\{f^{(j)}_{n+\varsigma}(x_i)>0\right\}}}.
\end{align*}
Averaging over five forecast horizons, we obtain measures of point forecast accuracies for all national and sub-national mortality series as
\begin{align*}
\text{Mean (MAFE)}  = \frac{1}{5}\sum^{5}_{h=1}\text{MAFE}(h), \quad \text{and} \quad
\text{Mean (RMSFE)} = \frac{1}{5}\sum^{5}_{h=1}\text{RMSFE}(h).
\end{align*}

\subsection{Point forecast results} \label{sec:5.2}

\begin{figure}[!htbp]
\centering
\includegraphics[width = 6.1in]{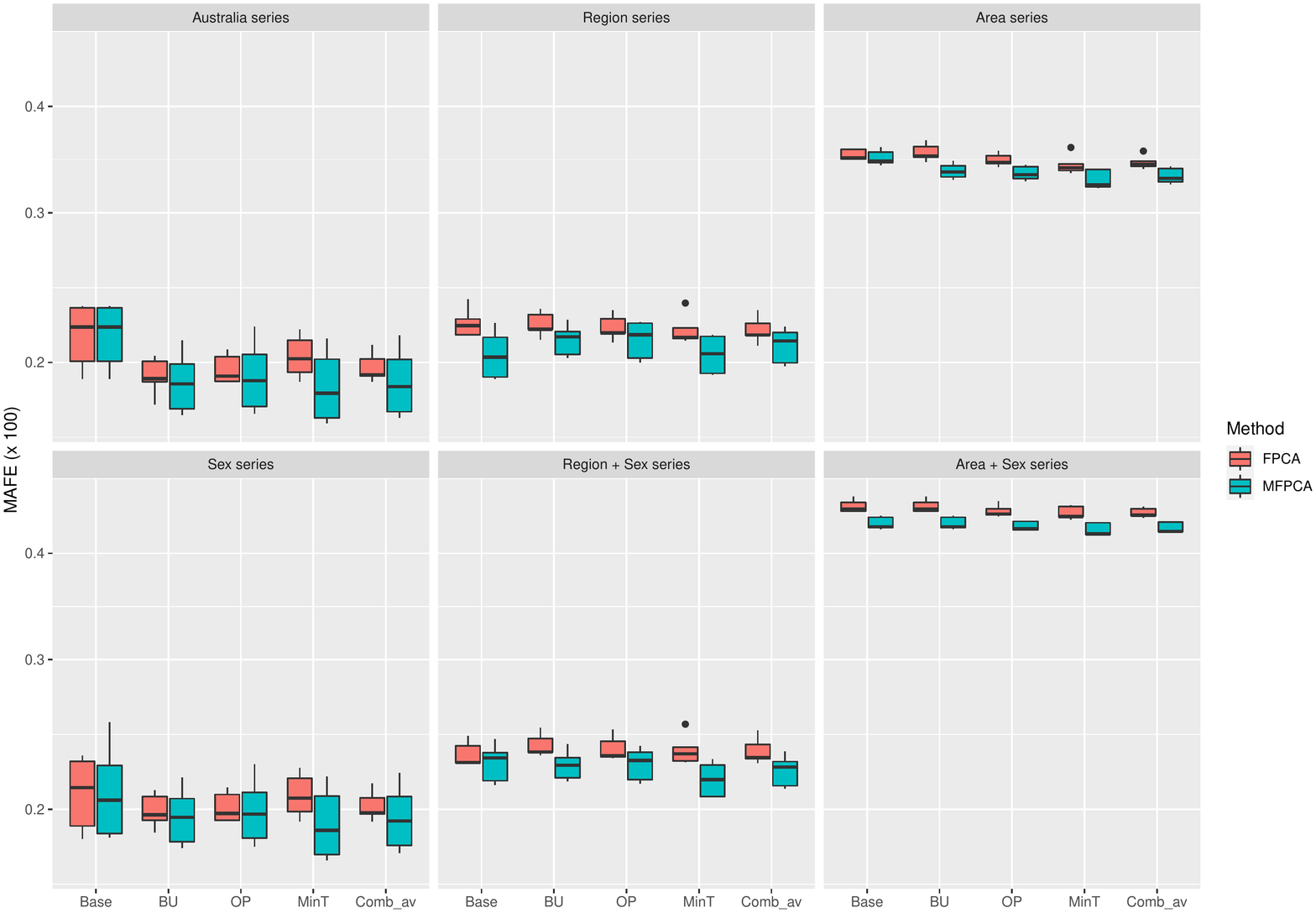}
\caption{MAFE $(\times 100)$ in the holdout sample between the univariate and multivariate functional time series methods.}
\label{fig:mafe}
\end{figure}

\begin{figure}[!htbp]
	\centering
\includegraphics[width = 6.1in]{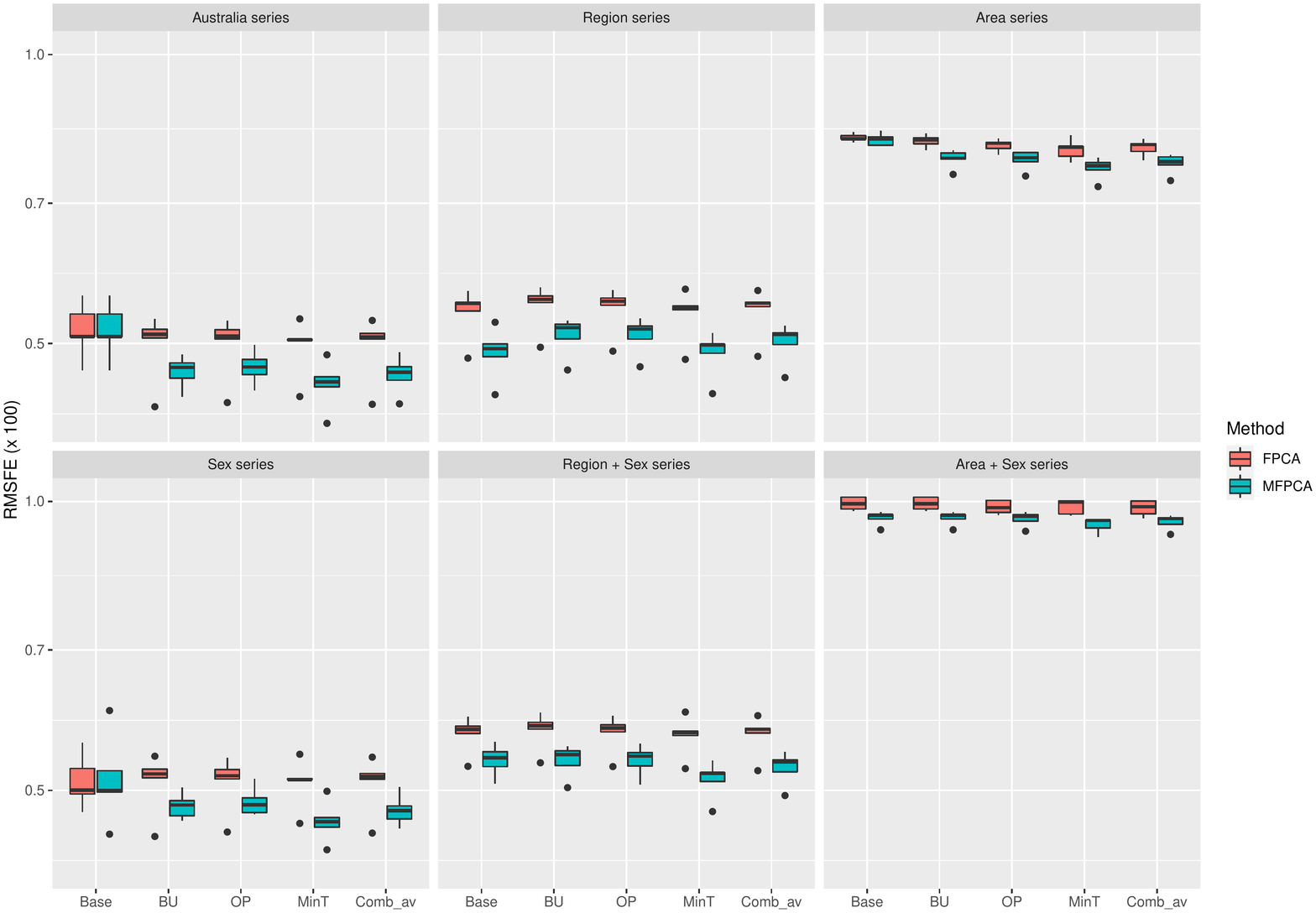}
	\caption{RMSFE $(\times 100)$ in the holdout sample between the univariate and multivariate functional time series methods.}
	\label{fig:rmse}
\end{figure}

We present the mean MAFEs ($\times 100$) for the base and reconciled forecasts in Figure~\ref{fig:mafe}, and the mean RMSFE ($\times 100$) in Figure~\ref{fig:rmse}, both by calculating the averages of all series at each disaggregation level of hierarchies considered. These forecasts are obtained by the proposed multivariate functional time series (MFPCA) method and its univariate (FPCA) counterpart. At all the sub-national levels, jointly modeling age-specific mortality rates produces more accurate point forecasts than considering each series individually using the FPCA method. The MFPCA method can extract common mortality features of strongly-related populations at a level of disaggregation. At the national level, where only one mortality time series exists, the FPCA model is equivalent to the MFPCA model with $\omega = 1$. (FPCA can be regarded as a special case of MFPCA when only a single series is considered.) Hence, both methods yield the same forecasts for the Australia series. 

Figures~\ref{fig:mafe} and \ref{fig:rmse} also show that pairing any of three reconciliation methods (BU, OP, and MinT), or using a mix of three (Comb\_av), with either MFPCA or FPCA improves the point forecast accuracy. Among the reconciliation methods considered, MinT yields the most accurate overall point forecasts when used with MFPCA. This result can be attributed to the MinT method incorporating all information from a full covariance matrix of forecast errors in obtaining a set of coherent forecasts. Both figures also indicate that forecast combination can help reducing point forecast errors, confirming findings of \cite{SH16} and \cite{Shang2020}.

\subsection{Robustness check} \label{sec:5.3}

\begin{figure}[!htbp]
\centering
\subfloat{\includegraphics[width = 6.3in]{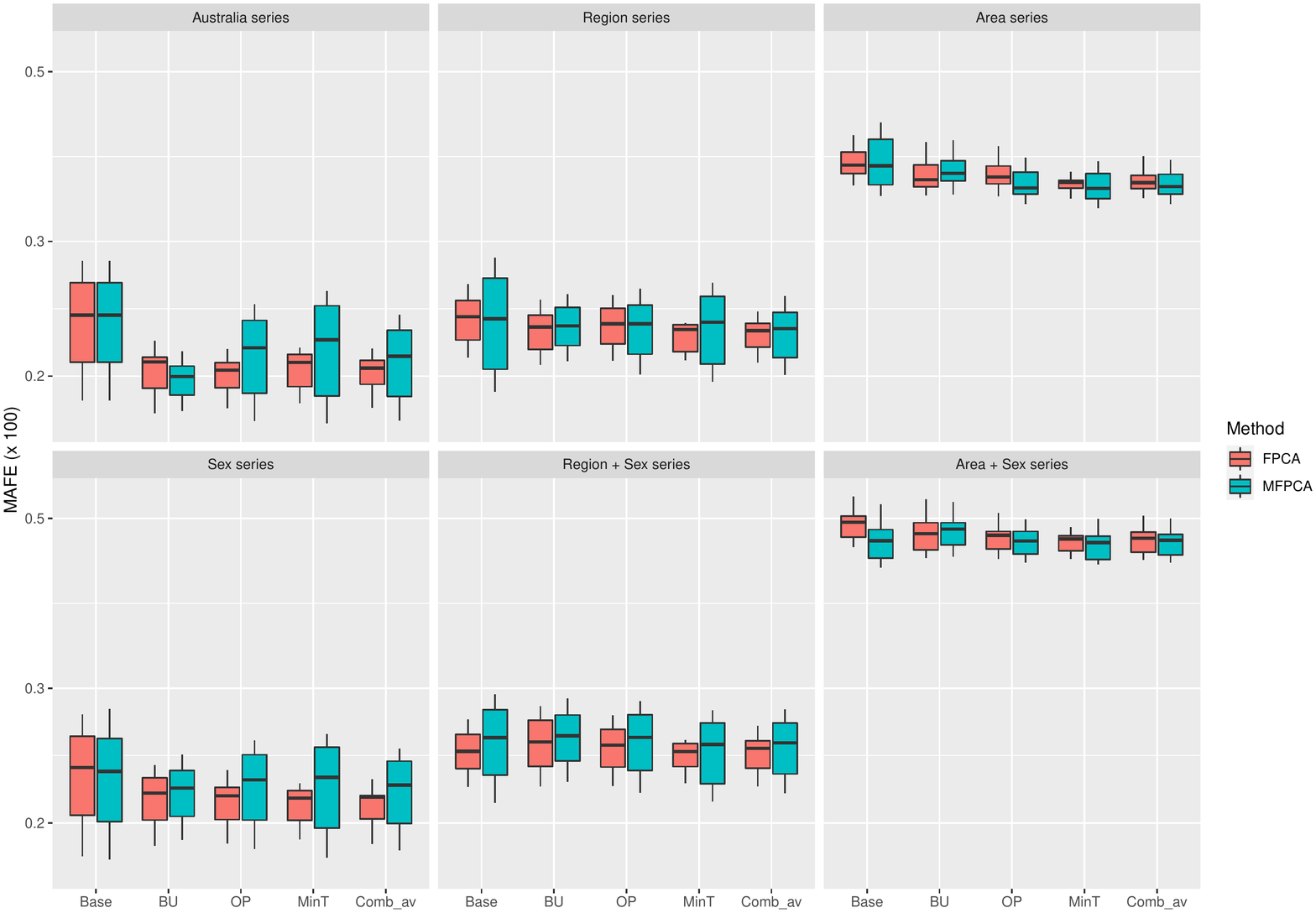}} 
\caption{MAFE ($\times 100$) in the holdout sample between the univariate and multivariate functional time series methods; $h = 1, \cdots, 10$.}
\label{fig:mafe_10}
\end{figure}

\begin{figure}[!htbp]
  \centering
  \subfloat{\includegraphics[width = 6.3in]{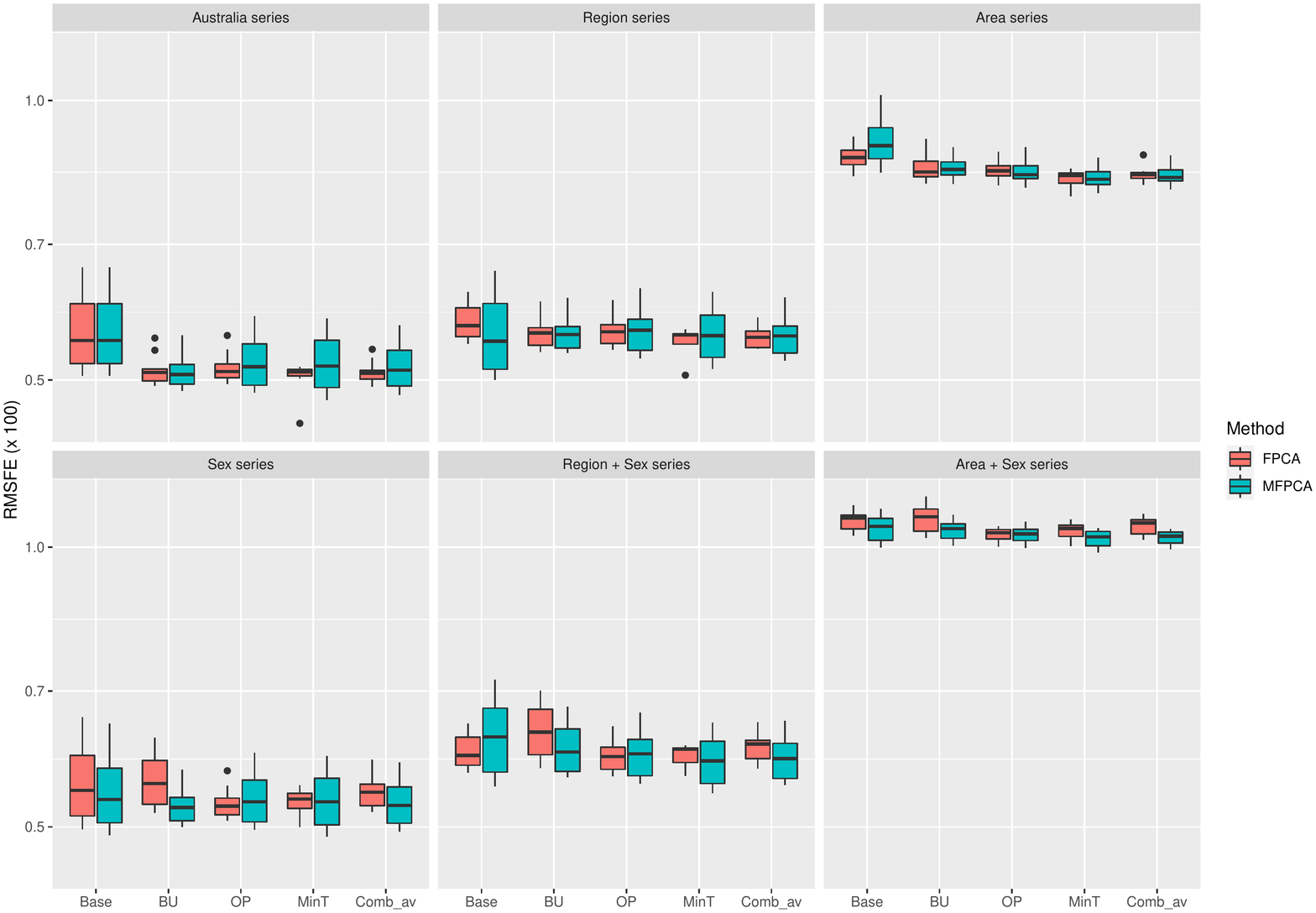}} 
  \caption{RMSFE ($\times 100$) in the holdout sample between the univariate and multivariate functional time series methods; $h = 1, \cdots, 10$.}
  \label{fig:rmse_10}
\end{figure}

We conduct an extended empirical application with one- to 10-step-ahead forecasts using the Australian age-specific mortality rates. We point out that the current dataset only covers death rates in regional and remote areas from 1993 to 2016. Moreover, most of the missing values in the dataset occur for the early years. Extending the forecasting horizon to $h=10$ produce less accurate forecasts than those in Section~\ref{sec:result}. We provide the mean MAFE and RMSFE results corresponding to the extended application in Figure~\ref{fig:mafe_10} nad Figure~\ref{fig:rmse_10}, respectively. It can be seen that the MFPCA method still outperforms the FPCA method in general.

\begin{figure}[!htbp]
\centering
\subfloat{\includegraphics[width = 6.3in]{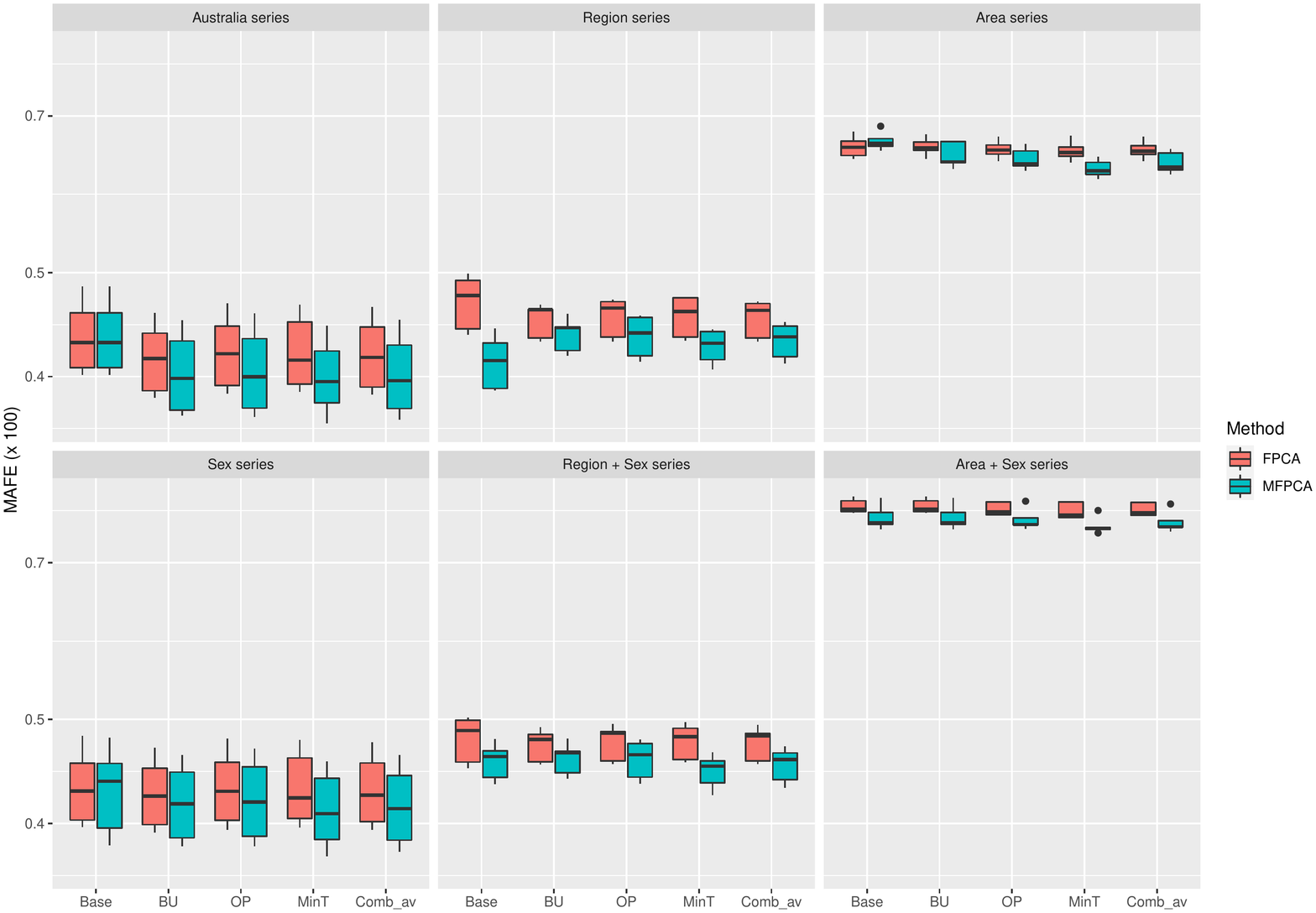}} 
\caption{MAFE $(\times 100)$ in the holdout sample between the univariate and multivariate functional time series methods, when mortality rates for ages between 50-95 are considered.}
\label{fig:mafe_rob}
\end{figure}

\begin{figure}[!htbp]
  \centering
  \subfloat{\includegraphics[width = 6.3in]{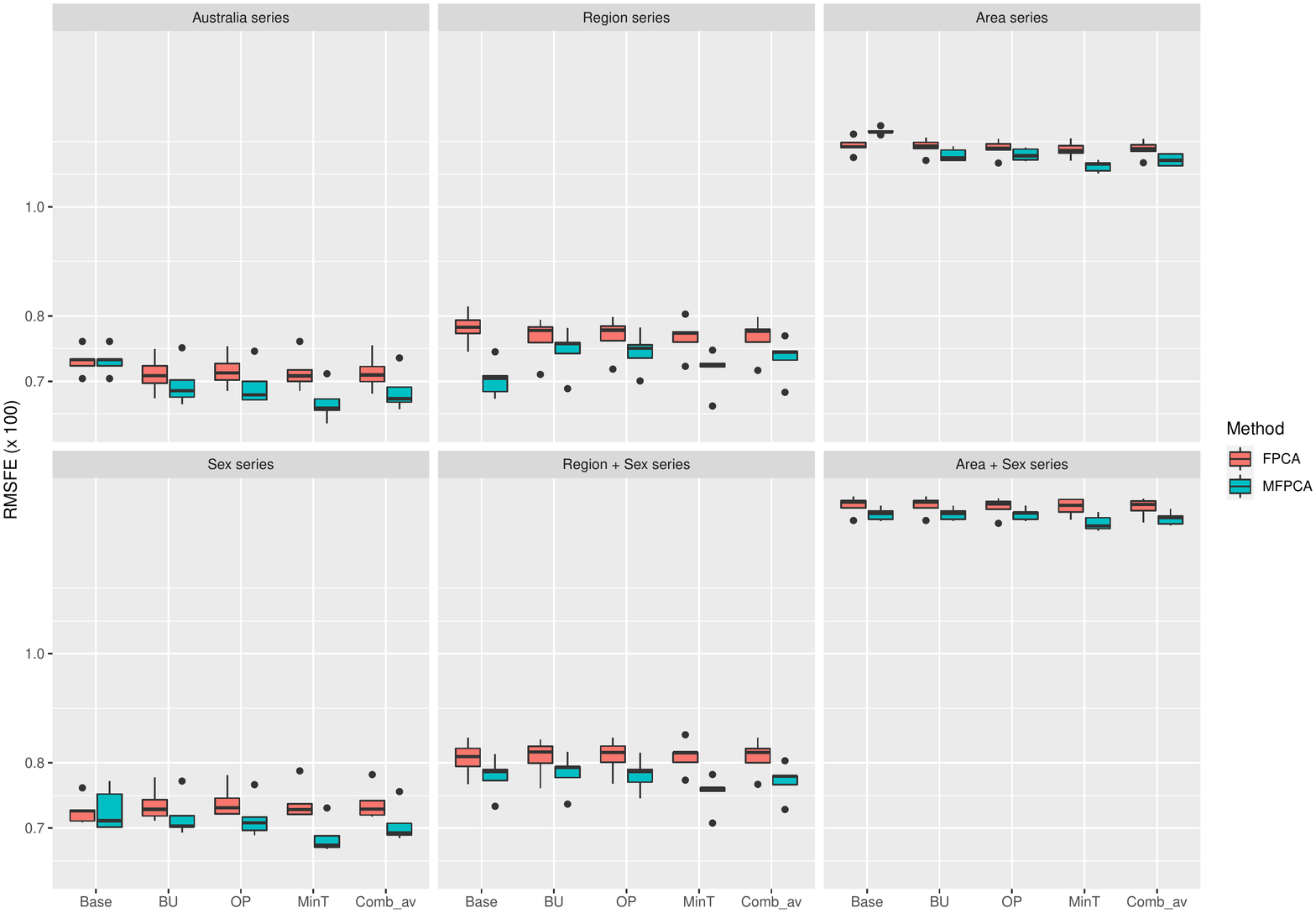}} 
  \caption{RMSFE $(\times 100)$ in the holdout sample between the univariate and multivariate functional time series methods, when mortality rates for ages between 50-95 are considered.}
  \label{fig:rmse_rob}
\end{figure}

Further, it is well known that mortality rates have different structures throughout ages. As shown in Figure~\ref{fig: image}, mortality rates for young people are generally more volatile than those for middle and old age. To check the robustness of our point forecasting results, we conduct an additional application using mortality rates for ages greater than 50. The mean MAFEs and RMSFEs of the robustness check are shown in Figure~\ref{fig:mafe_rob} and Figure~\ref{fig:rmse_rob}, respectively. We can see that the MFPCA method consistently outperforms the conventional FPCA method in forecasting age-specific mortality rates for ages between 50-85. Similar results reported in Sections~\ref{sec:5.2} and~\ref{sec:5.3} indicate the usefulness of the MFPCA method in modeling and forecasting age-specific mortality rates in a hierarchical structure.

\section{Conclusion} \label{sec:conclusion}

We extended the univariate functional time series forecasting method to a multivariate framework. The proposed multivariate functional time series method can be combined with reconciliation methods when applied to functional data formed by disaggregated series, such as sub-national age-specific mortality rates. The reconciliation methods, such as the bottom-up, optimal combination, trace minimization, and forecast combination methods, can improve point forecast accuracy. The complete approach described in this paper can be viewed as a grouped multivariate functional time series forecasting method.

The proposed method is applied to Australian age-specific mortality rates from 1993 to 2016. The out-of-sample forecasting results show the superior performance of our approach for point projections, compared with the conventional univariate functional forecasting method. Among the three forecast reconciliation methods, the trace minimization method shows the best performance in the study of Australian mortality data. A forecast combination method taking averages of the bottom-up, the optimal combination, and the trace minimization forecasts also has similar superior point forecast accuracy.

There are several ways in which the present paper can be further extended:
\begin{enumerate}
\item[1)] We considered two factors, namely sex, and geography, to disaggregate Australian sub-national mortality rates. It is possible to enrich the information contained in the group structure by differentiating these rates according to other factors, such as the cause of death \citep{ML97, GS15}, identification of Indigenous Australians (Aboriginal and Torres Strait Islander) \citep{Thomson1991, Supramaniam2006}, and native-born or immigrant status \citep{SGS01}. If appropriate data are available, we may attempt to extend the proposed grouped multivariate functional time series forecasting method to cause-specific mortality rates and Indigenous Australian mortality rates.
\item[2)] We have not considered the spatial dependence of Australian subnational populations. The spatial distance among populated areas as a covariate can be considered to incorporate the extent of correlation among populations, and thus build better disaggregation structures with homogeneous sub-groups. With an appropriate spatial distance-based clustering rule for sub-populations, we may develop more accurate mortality models and thus further improve the point forecast accuracy.
\item[3)] We only considered point forecasts in this study. It is also possible to compute interval forecasts to provide a precise estimate of the probability that the future realizations lie within a given range.
\item[4)] Optimal weights in forecast combination among the forecast reconciliation methods can also be a further pursuit to include more complicated algorithms such as cross-validation.
\end{enumerate}



\newpage
\bibliographystyle{agsm}
\bibliography{subnational}

\end{document}